\documentclass[aps,preprint,pre,epsf]{revtex4}
\usepackage{epsfig}
\begin{document}
\def\tauv{{\mbox{\boldmath{$\tau$}}}}
\def\sigmav{{\mbox{\boldmath{$\sigma$}}}}
\title{$1/d$-EXPANSION FOR $k$-CORE PERCOLATION}
\date{\today}
\author{A. B. Harris and J. M. Schwarz}

\affiliation{Department of Physics and Astronomy, University
of Pennsylvania, Philadelphia, PA, 19104}
\begin{abstract}

The physics of $k$-core percolation pertains to those systems whose
constituents require a minimum number of $k$ connections to each other in
order to participate in any clustering phenomenon. Examples of such a 
phenomenon range from 
orientational 
ordering in solid ortho-para ${\rm H}_2$ mixtures to the onset of rigidity in 
bar-joint networks to dynamical arrest in glass-forming liquids.  
Unlike ordinary ($k=1$) and biconnected ($k=2$) 
percolation, the mean field $k\ge 3$-core percolation transition is 
both continuous {\em and}
discontinuous, i.e. there is a jump in the order parameter accompanied
with a diverging length scale.  To determine whether or not this
hybrid transition survives in finite dimensions, we present a $1/d$ expansion
for $k$-core percolation on the $d$-dimensional hypercubic lattice.  We 
show that to order $1/d^3$ the singularity in the order
parameter and in the susceptibility occur at the same value of the
occupation probability.  This result suggests that the unusual hybrid nature of the
mean field $k$-core transition survives in high dimensions.

\end{abstract}
\maketitle

\section{INTRODUCTION}
In a number of physical problems long-range order requires
more than a single stranded path to propagate over long distances. 
One example of this is the propagation of quadrupolar order in
solid $(o-{\rm H}_2)_x(p-{\rm H}_2)_{1-x}$ (ortho-para H$_2$) mixtures~\cite{APM}. 
This system can be reasonably modeled as a quenched, site-diluted lattice of
electrostatic quadrupoles interacting via nearest-neighbor
interactions on an fcc lattice.
Since the lowest state of two such quadrupoles at displacement ${\bf r}$
is one in which one molecule may assume any orientation in the plane
perpendicular to ${\bf r}$, it is clear that long-range order can not
propagate down a long noncollinear single stranded path. Therefore, to
develop long-range order quadrupoles in an ``infinite cluster''
must have more than two nearest-neighboring quadrupoles.  

Another such example of multi-path long-range ordering is rigidity
percolation~\cite{FS,KW,FTG,DTT,TD} where each occupied site on a
lattice has $g$ degrees of freedom.  The degrees of freedom of the
site become fixed as more neighboring sites become occupied---one 
occupied neighbor constrains one degree of freedom. Therefore, 
in order to participate in the infinite rigid cluster, an occupied
site must have at least $g$ occupied neighbors.  Here, again, 
is an example of a constraint on the minimum number of 
occupied neighbors giving rise 
to multiple long-range paths through the system.  

More recently, an analogy between multi-path percolation and  
the onset of elasticity in repulsive soft spheres as 
the packing fraction of the system is increased 
has been put forth by Schwarz {\it et al}.~\cite{SLC}(SLC).  In the analogy overlaps between particles correspond to occupied 
neighboring sites.  To ensure local mechanical stability for each particle, 
$d+1$ occupied neighbors are required for each occupied site, otherwise the 
site is unstable and it is removed from the system (as opposed to running 
into other particles in the system). Here, the $d+1$ constraint 
gives rise to multiply connected paths 
that eventually span the system. The onset of elasticity in the repulsive soft sphere 
system---a type of jamming transition called Point J---is thought to have 
implications for other phenomena such as the glass transition and 
the colloidal glass transition~\cite{LN,OSLN}.
 
\begin{figure}
\centerline{\psfig{figure=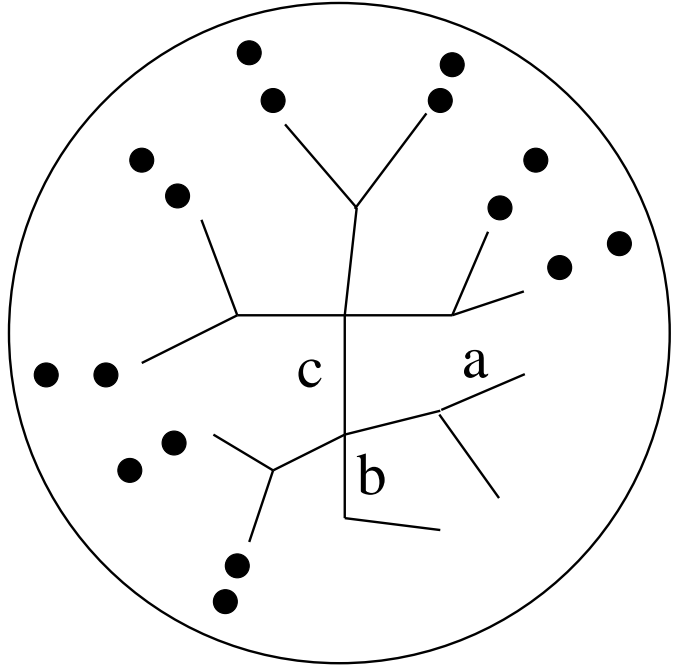}}
\vspace{0.3 in} \caption{Culling of a cluster for $k=3$ on a Bethe lattice
with coordination number $z=4$.  The filled circles
indicate branches which are $k-1$ connected to infinity.  Culling removes
the bonds at a, b, and eventually at c, including the lower two sets of 
filled circles, but the rest of the cluster survives.}
\label{CULL} \end{figure}

The particular model of percolation called $k$-core, or bootstrap,
percolation ~\cite{CLR,PSW,SLC} turns out to be the relevant model 
of interest for such systems. For both the solid ortho-para ${\rm H}_2$ mixtures and 
the jamming system, $k$-core percolation is an approximate description.  
However, for rigidity percolation, at least on the Bethe lattice, 
$g$-rigidity percolation is equivalent to $(g+1)$-core percolation~\cite{RIGIDCORE}.   
In $k$-core percolation, each
bond is independently occupied with probability $p$ and vacant with
probability $1-p$.  In addition, it is required that sites with less than $k$
occupied neighbors should be made vacant.  This "culling" operation
proceeds recursively until {\em all} remaining unculled occupied sites have
at least $k$ neighbors, as illustrated in Fig. \ref{CULL} for a 
Bethe lattice where each site has $z$ neighbors with $z=4$.  Such a model 
gives rise to ``many'' paths emanating to infinity from a single site for 
large enough $p$.  

Analysis of $k$-core percolation on the Bethe lattice by SLC 
showed that the critical point has some unusual
characteristics.  As was known previously~\cite{CLR}, they found that the 
transition at $p=p_0$---the $p$ at which an infinite cluster appears---is a
discontinuous one for $k\ge3$, and the probability $P_\infty$
that a site be in the infinite cluster (after culling) exhibits the
power law behavior:
\begin{eqnarray}
P_\infty&=& 0, \hspace{1.4 in} p < p_0 , \nonumber \\
P_\infty  &=& P + a(p-p_0)^{1/2}\ \ \ \ \ \  p \ge p_0 \  .
\label{EOSEQ} \end{eqnarray}
More strikingly, SLC found a diverging correlation length and that 
an appropriate susceptibility has a power law divergence at the
same threshold value of $p$ at which the order parameter has a
discontinuity.  

An immediate question then arises, is this
unusual structure of the critical point characteristic of infinite
$d$, or does it survive for finite $d$?  A clear answer to this
question could be provided if a field theoretical formulation of
the problem were available, but at present, no such field theory
exists. (Although a field theory for the jamming transition 
based on force balance, as opposed to local mechanical stability, 
 has been recently proposed~\cite{BRANDEIS},  it
is not obvious that this model is in the same universality class 
as $k$-core percolation in finite dimensions.)
As for lattice models in low $d$ spatial dimensions, 
until recently, numerical studies indicate that the $k$-core transition is 
{\em either} continuous or $p_c=1$, i.e. nothing survives the culling process until 
the lattice is fully occupied~\cite{KL,AL}. A case of the former is $k=3$ on the triangular lattice~\cite{MC}, and 
for the latter, $k=3$ on the square lattice~\cite{KL}. In fact, for $k=3$ on the 
square lattice    
it has been proven by van Enter~\cite{ACDvanE} that $p_c=1$~\cite{DAWSON}.  Moreover, it has recently been proven for hypercubic lattices 
that $p_c=1$ for $k\ge d+1$~\cite{TBF}.  To date, the only 
 numerical evidence for a $k$-core transition with a jump in the 
order parameter and yet a diverging correlation length is a {\em directed} 
$k$-core model with 10 nearest neighbors in two dimensions 
studied by SLC~\cite{privcommBIROLI}.  This result supports the notion that 
the unusual nature of the $k$-core transition found in the mean field 
survives in finite dimensions, though simple lattices, like the square 
lattice, are probably not the place to look for such a transition.  

Lacking a $k$-core field theory and the sparsity of finite $d$ 
numerical results, 
we decided to implement an expansion in
$1/d$.  As one sees for the Ising model, this expansion 
cannot be used to discuss critical exponents, because these
exponents are independent of $d$ for $d>d_c$, where $d_c$ is the
so-called upper critical dimension. For the Ising model, 
$d_c=4$~\cite{Wilson,Fisher} and for ordinary percolation is
$d_c=6$~\cite{PT,ABHTCL}.  However, this expansion has been used to
generate short series in $1/d$ for the critical value of the coupling
constant in problems such as the Ising model and self-avoiding
walks,~\cite{ISING} for spin glasses,~\cite{SINGH} for lattice
animals,~\cite{ABH} and for ordinary percolation~\cite{HDPERC}.
Thus, this expansion is ideally suited to answer the question or 
whether or not the unusual $k$-core mean field transition survives in 
finite dimensions.  It should also be noted that the interpolation to
continuous dimension implied by this expansion is precisely the
same~\cite{BB} as that used by Wilson in his development of the renormalization
group~\cite{Wilson}.  We will use this expansion to calculate corrections 
up to order $d^{-3}$ for the critical coupling constant at which 
the discontinuity
in the order parameter takes place and compare it to the critical
value at which
the susceptibility diverges.  Since we find that these two threshold
remain equal up to this order in $1/d$, we conclude that it is likely
that this coincidence is robust and remains true at least for some
range of high $d$.  We note that these results 
are stronger than the results of Toninelli {\it et al}.~\cite{TBF} showing that the 
unusual nature of the transition survives on Husimi trees (a Bethe lattice 
with finite loops), since, like the Bethe lattice, this structure
has an infinite fractal dimension and therefore provides no information
on the situation for finite $d$.

Briefly this paper is organized as follows.  In Section II we review
known results for the Bethe lattice to fix our notation. Section III 
introduces the notion of perturbing the equation of state.  Section IV 
introduces the concepts behind the $1/d$ expansion. Section V 
presents the perturbative $1/d$ corrections to the equation of state, 
while Section VI does the same for the susceptibility.  In Sec. VII
we summarize the conclusions which may be drawn from our result,
that to order $1/d^3$ at large $d$, the $k$-core transition remains 
a hybrid transition and in Sec. VIII we discuss implications of our result 
for systems such as glass-forming liquids.  

\section{BETHE LATTICE EQUATION OF STATE}

\begin{figure}
\centerline{\psfig{figure=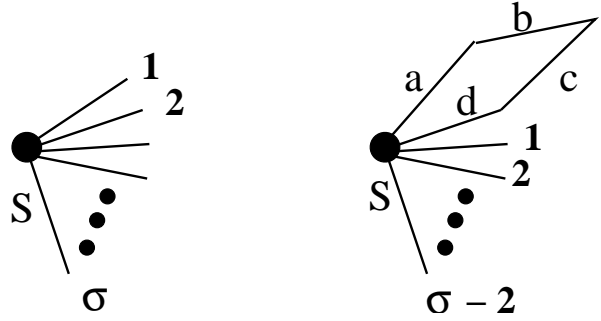}}
\vspace{0.3 in}
\caption{Diagrammatic representation of the EOS.  Left:
the unperturbed EOS based on a seed site, S, with $\sigma$ emerging bonds.
Right: the perturbation to the EOS from the insertion of a square, so that
now the seed site has $\sigma -2$ single bonds and an attached square.}
\label{SQ7} \end{figure}

In this section we construct the self-consistent equation for the order
parameter on the Bethe lattice for $k$-core bond percolation. For this
purpose we consider a rooted Bethe lattice, i.e. a lattice emanating
from a seed site (as in Fig. \ref{SQ7}) in which the lattice is constructed
by recursively adding $\sigma \equiv z-1$ sites to each bond.
To construct the self-consistent equation, the missing $z$th neighbor of
the seed site is posited to survive culling.  Therefore, the entire
cluster will survive culling if each recursively added site has
$k-1$ outward bond connections to infinity.  We therefore define the
quantity $P_\infty$ to be the probability that when we add a site
to the cluster, that site is then has $k-1$ outward bond connections
to infinity which survive culling.
The probability that some site on a Bethe lattice is in the $k$-core
can then be related to and has the same type of singular behavior
as $P_\infty$ for $k \geq 3$~\cite{SLC}. Accordingly, we have the
self-consistent equation of $P_\infty$ as~\cite{CLR,SLC}
\begin{eqnarray}
P_\infty &=& 1 - \sum_{m=0}^{m=k-2} {\sigma! \over (\sigma-m)!m!}
(1 - p P_\infty)^{\sigma-m} (pP_\infty)^m\,,
\end{eqnarray}
which we write in terms of $Q \equiv pP_\infty$ as
\begin{eqnarray}
Q/p &=& 1 - \sum_{m=0}^{m=k-2} {\sigma! \over (\sigma-m)!m!}
Q^m (1 - Q)^{\sigma-m} \equiv \Phi(Q)\,.
\label{EQKA} \end{eqnarray}
Starting at $p=1$, so that $Q=1$, we consider the effect of reducing
$p$ and find that
\begin{eqnarray}
{dQ \over dp} &=& (Q/p) \left[ {1 \over p} - {d\Phi \over dQ} \right]^{-1}\,.
\end{eqnarray}
As $p$ is decreased, $Q$ decreases until $p$ reaches a critical value
$p_c$ at which
\begin{eqnarray}
{1 \over p} &=& {d\Phi(Q) \over dQ}
= {(\sigma-k+1)! \over (\sigma-k+1)! (k-3)!} Q^{k-2} (1-Q)^{\sigma-k+1}\,. 
\label{EQKB} \end{eqnarray}
Indeed, when the description of the transition is phrased in
this way, it seems almost obvious that "turning on"
finite spatial dimension will not invalidate Eq. (\ref{EOSEQ}).
 
\section{PERTURBATIVE CORRECTIONS TO $\Phi(Q)$}

Now we wish to perturb the equation of state (EOS).    For instance, we may consider what
terms appear in the expansion of the EOS which depend
on the variables $p_ap_bp_cp_d$, where $a$, $b$, $c$, and $d$ form a square.
We will construct
\begin{eqnarray}
\Delta \Phi = \Phi_H - \Phi_B \ ,
\label{INSERTEQ} \end{eqnarray}
where the subscript on $\Phi$ indicates whether it is to be
evaluated for the hypercubic (H) lattice or the Bethe (B) lattice.
This expression includes the additional iterative term (which we only
invoke once at order $1/\sigma^2$ where $\sigma$ is $\mathcal{O}(d)$) for the H lattice and it also
takes account of terms which appear on the B lattice but which do
not have counterparts on the H lattice.

To evaluate the derivative with respect to the $p$'s in this expansion, we note that
each  occupied bond carries a factor of $p$ and each unoccupied
bond carries a factor of $1-p$.  So for any bare diagram of 
$b$ bonds, the derivative with respect to all its $p$'s will involve
a sum of the $2^b$ configurations of occupied and unoccupied bonds,
in which an occupied bond carries a factor $+1$ and an unoccupied
bond carries a factor $-1$.  Thus, from a square (which is the
configuration giving the leading correction at relative order
$1/\sigma^2$) we generate 16 subdiagrams (the first of which
is just the one for which all bonds are occupied and the last of
which is the one for which all bonds are unoccupied).
\begin{figure}
\centerline{\psfig{figure=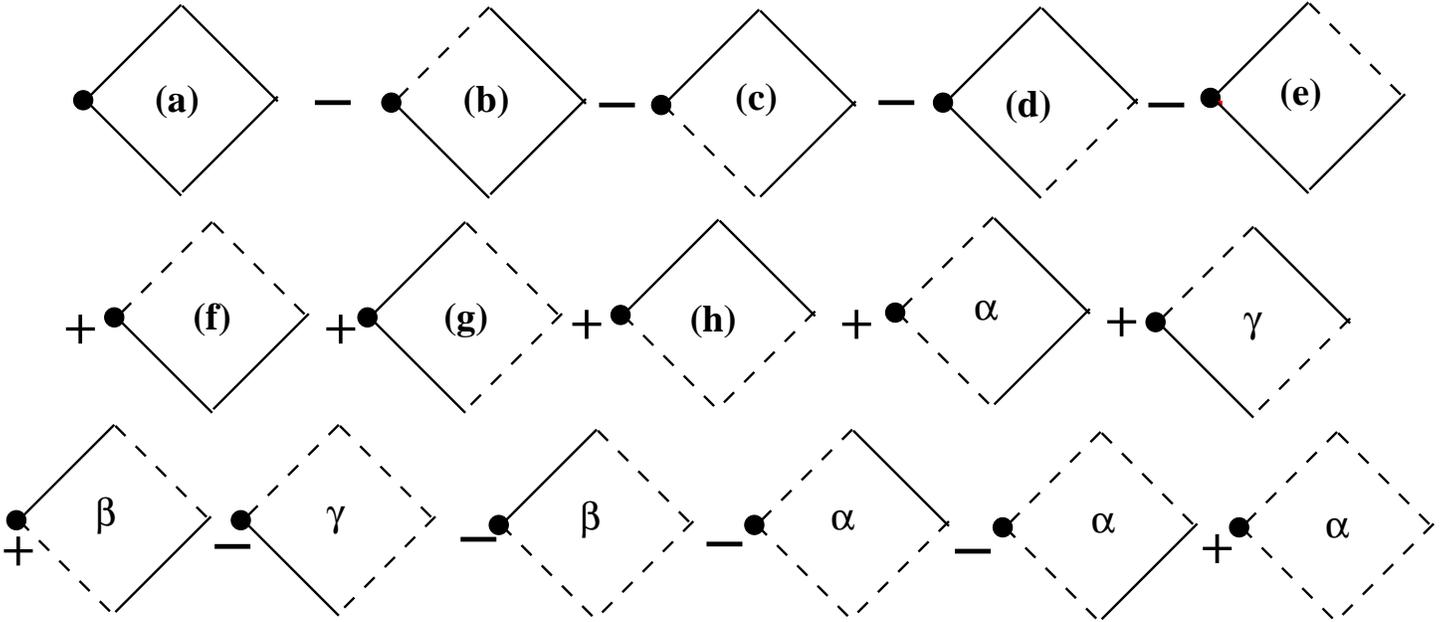}}
\vspace{0.3 in} \caption{Contributions from the 16 subdiagrams (of
diagram a) which result from either occupying (full line) or not
occupying (dashed line) each bond of the square on the H lattice.
Greek letters label sets of subdiagrams which cancel one another.}
\label{F13} \end{figure}
\begin{figure}
\centerline{\psfig{figure=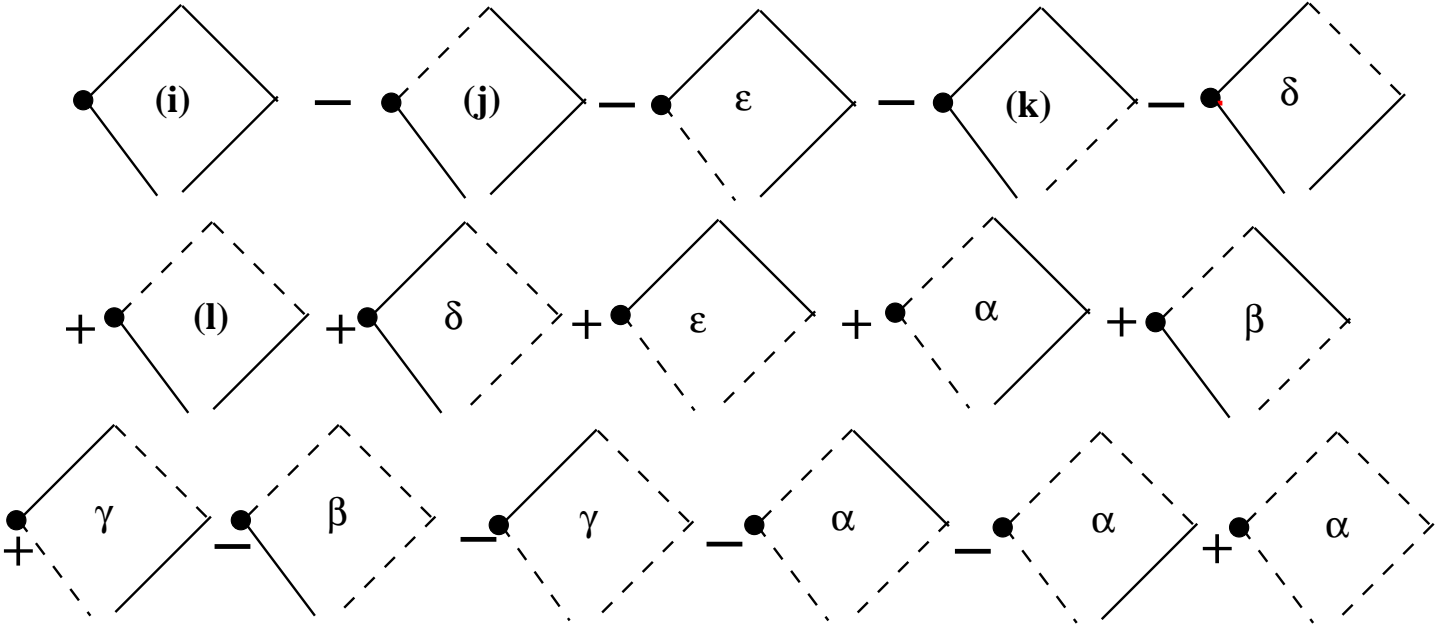}}
\vspace{0.3 in} \caption{As above, but for the B lattice. Vertices
which are close to one another would coincide on the H lattice.}
\label{F14} \end{figure}
Accordingly, the derivative with respect to the four $p$'s which
form a square (the leading correction to the B lattice) gives
rise to Figs. 1 for the H lattice and to Figs. 2-5 for the
B lattice.

\begin{figure}
\centerline{\psfig{figure=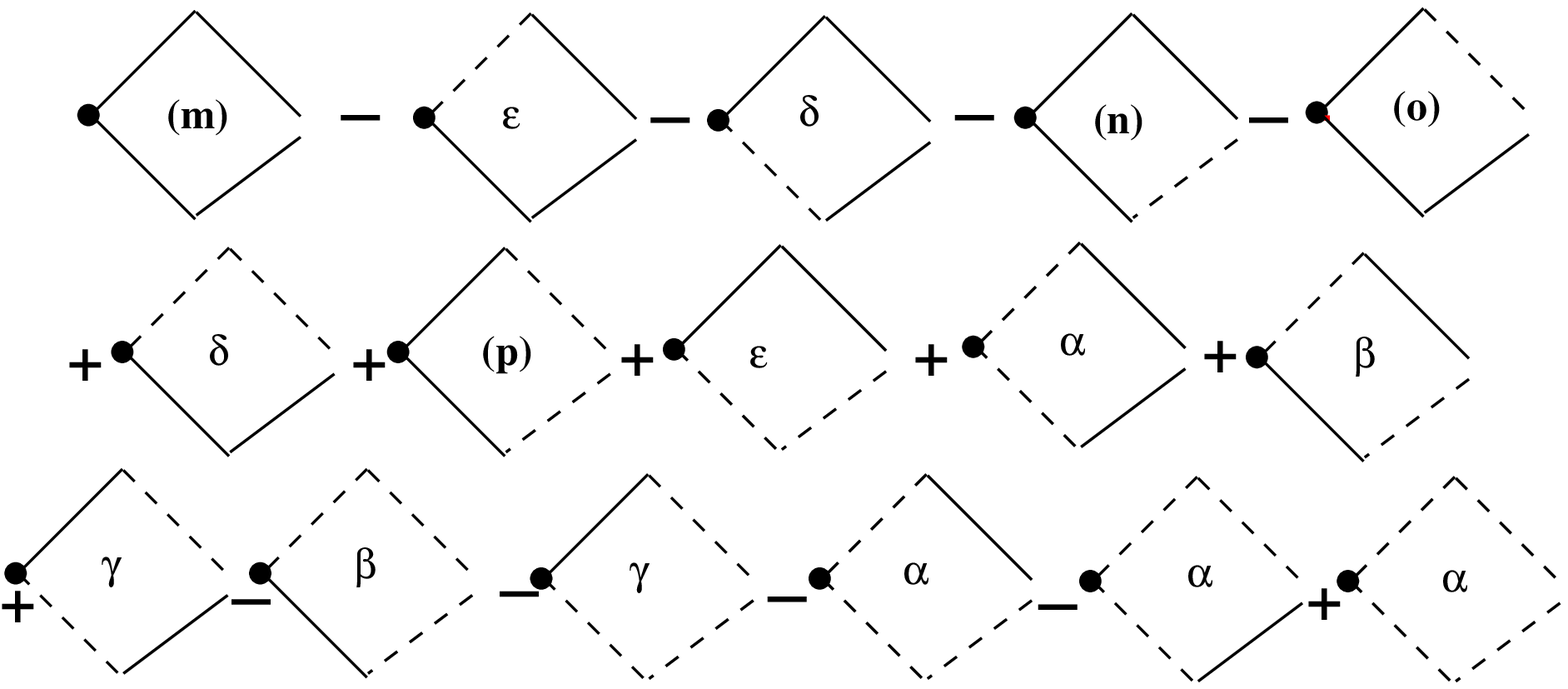}}
\vspace{0.3 in} \caption{As above, but for the B lattice.}
\label{F15} \end{figure}

\begin{figure}
\centerline{\psfig{figure=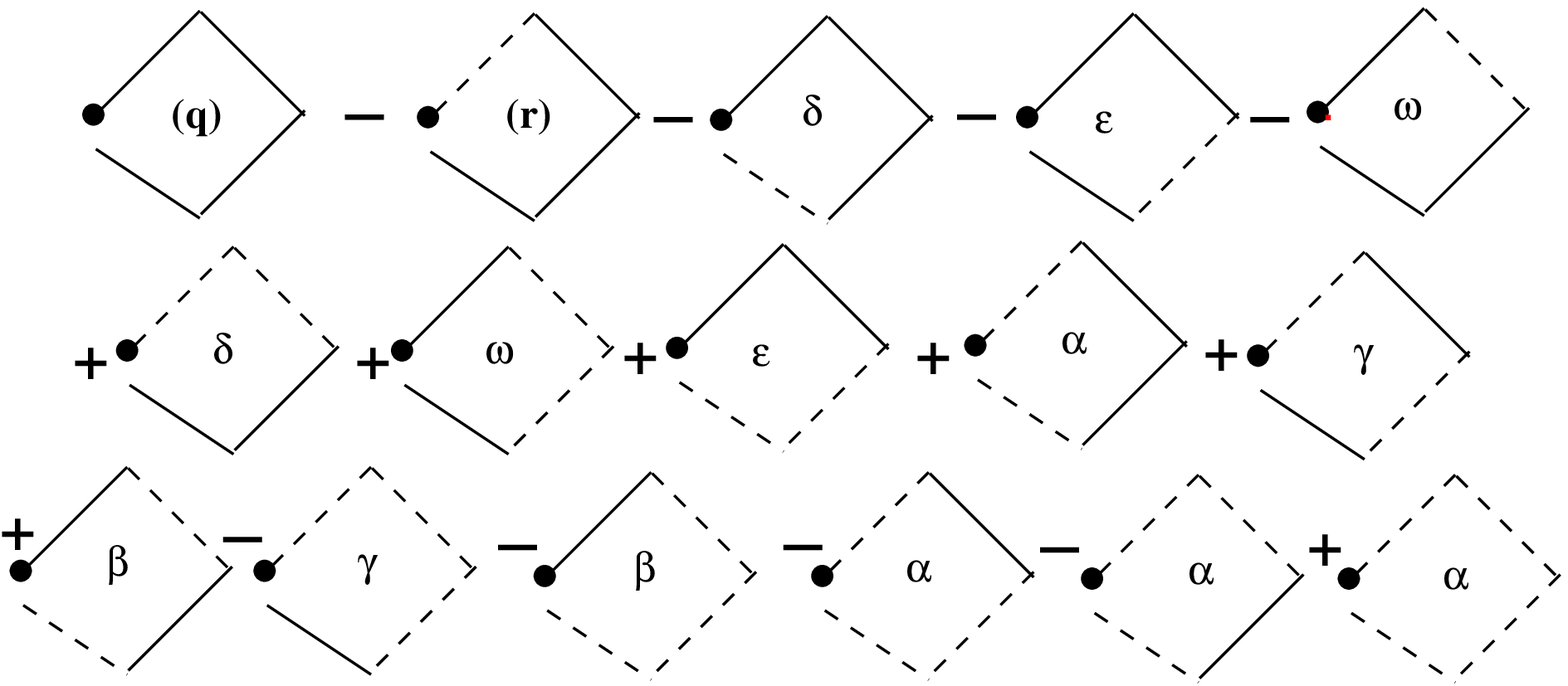}}
\vspace{0.3 in} \caption{As above, but for the B lattice.}
\label{F16} \end{figure}
If $f(x)$ denotes the contribution to $\Delta \Phi$ from square
insertions, which we call $\Phi_{\rm Sq}$, then 
\begin{eqnarray}
\Phi_{\rm Sq} &=& f(a) - 2f(b) - 2 f(d) - 2 f(i)
+2 f(j) +2 f(k) -f(m) + 2 f(n) -2f(q) + 2 f(r) \ ,
\label{SQPHI} \end{eqnarray}
\begin{figure}
\centerline{\psfig{figure=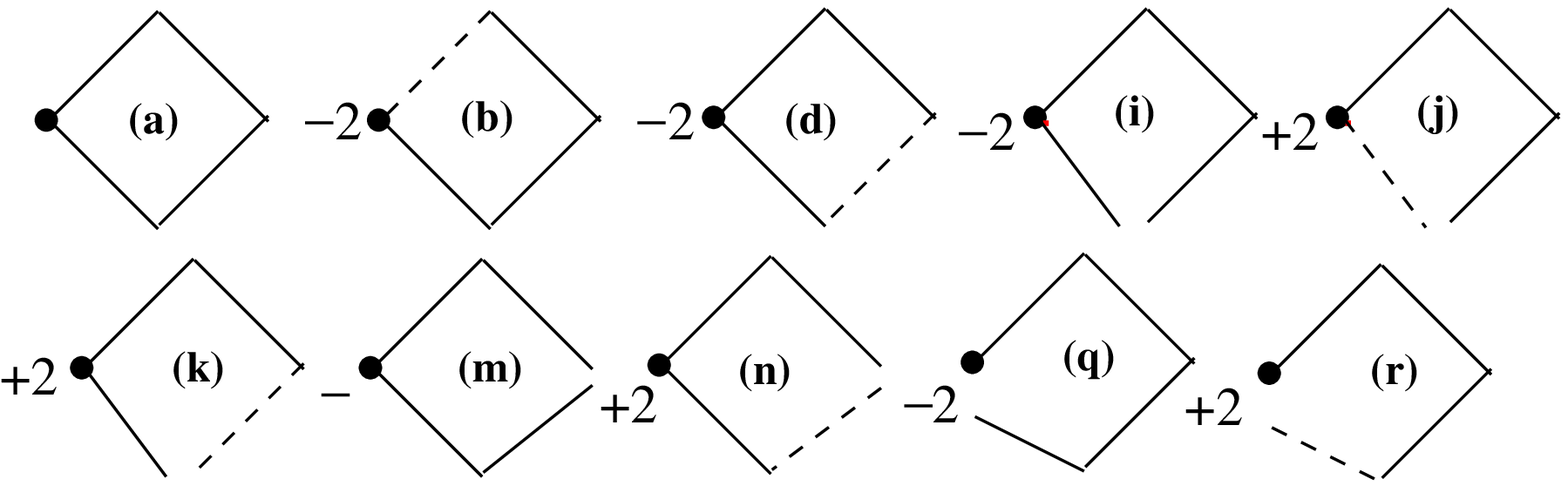}}
\vspace{0.3 in} \caption{Diagrammatic representation of Eq. \ref{SQPHI}.  Note that there are no diagrams with two unoccupied bonds.  This is because 
for every H lattice diagram there is a B lattice equivalent with a ``cut'' 
between the two unoccupied bonds resulting in a cancellation.}
\label{DIAEQ7} \end{figure}

\noindent
where we noted that $f(f)=f(h)=f(l)$ and $f(g)=f(p)$.  This
equation is represented in Fig. \ref{DIAEQ7}.  
The factors of 2 take account of the fact that some topologies occur
in two equivalent realizations.  When we calculate the contributions $f$
we must include not only the factor $p_ap_bp_cp_d=p^4$, but also the number of
ways (which depends on  $\sigma$) the bonds making up the square
can be selected out of the $\sigma$ available bonds.
Since this factor is of order $\sigma^2$, we see that $\Phi_{\rm Sq}$
is of order $p^4 \sigma^2$ which we will see is of order $\sigma^{-2}$.  
As will also be seen later, 
if we were only evaluating these corrections to leading
order in $1/\sigma$, then we would not need to differentiate between diagrams
(b), (j), and (r), or between (d), (k), and (n).  Accordingly, at order $1/\sigma^3$,
when we consider the analogous contribution from hexagons, $\Phi_{\rm Hex}$,
we do not have to be explicit about the configuration of the vacant bond(s)
and we therefore have the hexagon insertions shown in Fig. \ref{H1a}.

\vspace{0.3 in}
\begin{figure}
\centerline{\psfig{figure=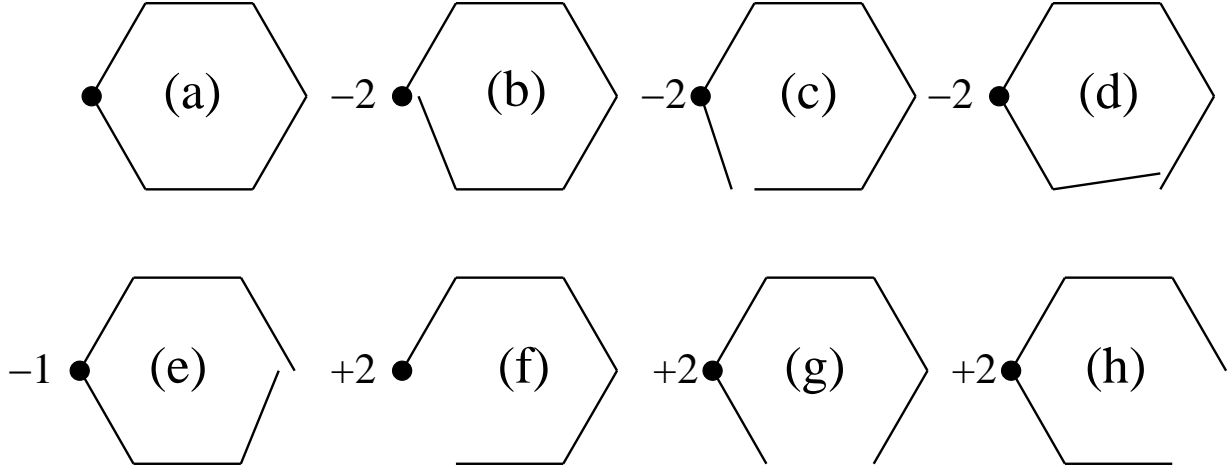}}
\vspace{0.3 in} \caption{As in Fig. \ref{DIAEQ7}, contributions to $\Phi(Q)$ from hexagons
on either a H or a B lattice. Diagrams (f), (g) and (h) correspond to 
sums of graphs on both the H and B lattice.  For example, the square 
equivalent to hexagon  
diagram (f) is the sum of diagrams (b), (r), and (j) in 
Fig. \ref{DIAEQ7}.    Diagrams similar to these give the
contribution to the susceptibility, $\Gamma_{H,A}$, due to hexagon
insertions as we will eventually show.}
\label{H1a} \end{figure}

\subsection{Percolation}

We can apply the above formulation to calculate the corrections to the
critical concentration $p_c$ in the ordinary bond percolation problem, and
check the results by comparing to those obtained by Gaunt and Ruskin~\cite{HDPERC}. 
For percolation, a diagram contributes to the probability of the seed
being connected to infinity if one or more of its vertices is connected
to infinity.  So, for the rooted Bethe lattice the EOS at the transition is
\begin{eqnarray}
P &=& 1 - (1-pP)^\sigma \ .
\end{eqnarray}
(The only nonpercolating case is if all vertices are {\it not}
connected to infinity.)  Since the percolation transition is
continuous, the critical concentration is found by expanding the
EOS in powers of $P$.  At linear order the solution requires that
\begin{eqnarray}
1 &=& \sigma p_c \ .
\end{eqnarray}
For hypercubic lattices the contribution to $\Phi_{\rm Sq}$ from
Eq.  (\ref{SQPHI}) is written as
\begin{eqnarray}
\Phi_{\rm Sq}={1 \over 2} (\sigma -1)^2 p^4 \Biggl[[1-(1-pP)^{\sigma-2}(1-P^{\sigma-1})^3] -2[1-(1-pP)^{\sigma-2}(1-P^{\sigma-1})^3] \nonumber\\  
-2[1-(1-pP)^{\sigma-2}(1-P^{\sigma-1})^3] 
-2[1-(1-P)^2(1-pP)^{\sigma-2}(1-P^{\sigma-1})^2] \nonumber\\
+2[1-(1-P)(1-pP)^{\sigma-2}(1-P^{\sigma-1})^2]
+2[1-(1-P)(1-pP)^{\sigma-2}(1-P^{\sigma-1})^2]\nonumber\\
-[1-(1-P)^2(1-pP)^{\sigma-2}(1-P^{\sigma-1})^2]
+2[1-(1-P)(1-pP)^{\sigma-2}(1-P^{\sigma-1})^2] \nonumber\\
-2[1-(1-P)(1-pP)^{\sigma-1}(1-P^{\sigma-1})^3]
+2[1-(1-pP)^{\sigma-1}(1-P^{\sigma-1})^3] \Biggr], 
\end{eqnarray}
where the terms between each small square bracket denote diagrams (a), (b), (d), etc. from Fig. \ref{DIAEQ7}, and
\begin{eqnarray}
P^{\sigma-1} &=& 1 - (1-pP)^{\sigma-1} + {\cal O}(\sigma^{-2}). 
\end{eqnarray}
$P^{\sigma-1}$ is the probability that a site has at least $k-1$ connections 
to infinity out of $\sigma-1$ bonds attached to the site.  A generalization 
of this quantity will be introduced later on. Then $\Phi_{\rm Sq}$ 
can be simplified to read
\begin{eqnarray}
\Phi_{\rm Sq}={1 \over 2} (\sigma -1)^2 p^4 \Biggl[ -2P(1-pP)^{\sigma-1}(1-P^{\sigma-1})^3 + 3(1-pP)^{\sigma-2}(1-P^{\sigma-1})^3 \nonumber\\
+ 3(1-pP)^{\sigma-2}(1-P^{\sigma-1})^2(1-P)^2 - 6(1-pP)^{\sigma-2}(1-P^{\sigma-1})^2(1-P)\Biggr].
\end{eqnarray}
\vspace{0.3 in}
\begin{figure}
\centerline{\psfig{figure=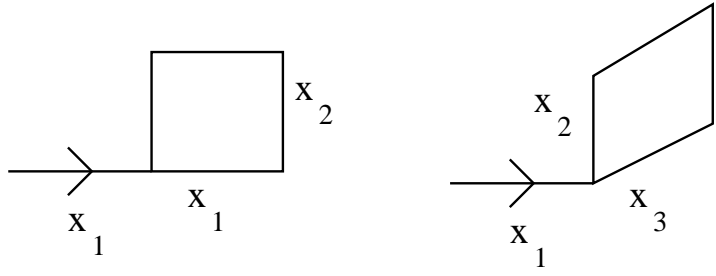}}
\vspace{0.3 in} \caption{Attaching a square to a site.}
\label{N} \end{figure}
The factors of $(1-pP)^{\sigma-2}$ and $(1-pP)^{\sigma-1}$ ensure that
the bonds attached to the seed site but not in the square are isolated from
infinity, and the factor $N_{\rm Sq} = (\sigma-1)^2/2$ counts the number
of possible squares.  We construct  $N_{\rm Sq}$ using Fig. \ref{N}
as follows.  Say the bond to which we wish to attach the square is
in the $x_1$ direction.  Then there are two cases: either the square
involves a bond in $x_1$ direction or it doesn't.  In the first case
the square involves the $x_1$ bond and one of $2d-2=\sigma-1$
choices for the other axis of the square.  The $x_1$ bond can either
be the first or the second leg of the square, but we ignore this
degeneracy because it simply corresponds to traversing the square
in different senses.  So for this case
there are $(\sigma-1)$ configurations.  In the second case
the square involves two  out of $d-1$ dimensions, so there are
$4(d-1)(d-2)/2=(\sigma-1)(\sigma-3)/2$ such configurations.  In all
\begin{eqnarray}
N_{\rm Sq}= (\sigma-1)^2/2\ .
\label{NSQEQ} \end{eqnarray}
This number would be twice as large if
the sense in which the square was traversed mattered.
For hexagons the calculation is much simpler because we do not need to
keep track of the configuration of vacant bonds, which forces us
to distinguish between $P$ and $P^{\sigma-1}$ and which leads to
corrections of  relative order $1/\sigma$. Using Fig. \ref{H1a},
we write the contribution to $\Delta \Phi$ from hexagons in leading
order of $1/\sigma$ as
\begin{eqnarray}
\Delta \Phi_{\rm Hex} &=& - 2\sigma^3 p^6 (1-pP)^{\sigma-2}
\Biggl[ 5 (1-P)^5 \left( 1 - [1-P] \right)  + 2 (1-pP)(1-P)^5 \left(
1 - [1-P] \right)  \Biggr]
\nonumber \\ &=& - 2 \sigma^3 p^6 (1-pP)^{\sigma-2} P(1-P)^5
[5 + 2(1-pP)] \ . 
\end{eqnarray}
Here the factor $2 \sigma^3$ is (to leading order in $1/\sigma$) the
number of ways of attaching a hexagon to a site.
Note that $p \rightarrow 1/\sigma + {\cal O}(\sigma^{-3})$ 
so that to the order we need, $P^{\sigma-1} \sim P [1 - (1/\sigma)]$.  Also 
note that for this continuous transition we only need the contribution
to $\Delta \Phi$ which are linear in $P$.  Thus we set
$\Delta \Phi_{\rm Hex}=-14P \sigma^{-3}$ and then
\begin{eqnarray}
\Delta \Phi &=& - {P \over 2\sigma^2 } \left[ 1 - {1 \over \sigma} \right]^2 
[2 + 3[1 - (1/\sigma)] -  14 {P \over \sigma^3} \ .   
\end{eqnarray}
Finally, using $1=\sigma p_c + d\Delta \Phi /dP$, we get
\begin{eqnarray}
\sigma p_c &=& 1 + (5/2) \sigma^{-2} + (15/2) \sigma^{-3} + {\cal O} (\sigma^{-4}) \ ,
\label{GRPERC} \end{eqnarray}
in agreement with Ref. \onlinecite{HDPERC}.

\section{EXPANSION IN POWERS OF $1/\sigma$}

Before getting into the calculation we should indicate how the
various variables are to be expanded in powers of $1/\sigma$.
Consider Eqs. (\ref{EQKA}) and (\ref{EQKB}).  Set
\begin{eqnarray}
p_0 &=& \sum_n \beta_n \sigma^{-n} \ , \ \ \
Q_0 = \sum_n \alpha_n \sigma^{-n} \ ,
\label{PZERO} \end{eqnarray}
where the subscripts "0" indicate that the quantities are those of the
Bethe lattice solution. Although we will not invoke the actual values
of the coefficients in these expansions, we will, for illustrative
purposes, determine the leading coefficients, $\alpha_1 \equiv \alpha$
and $\beta_1 \equiv \beta$, where 
\begin{eqnarray}
{\alpha \over \beta} &=& 1 - e^{-\alpha} \sum_{m=0}^{m=k-2}
{\alpha^m \over m!} 
\end{eqnarray}
and
\begin{eqnarray}
1 &=& \beta \alpha^{k-2} e^{-\alpha } \ .
\end{eqnarray}
By eliminating $\beta$, we obtain an equation which determines
$\alpha$:
\begin{eqnarray}
e^\alpha = \alpha^{k-1} + \sum_{m=0}^{m=k-2} {\alpha^m \over m!} \ .
\end{eqnarray}
For $k=3$, numerical evaluation yields the approximate values
$\alpha=1.793282133$ and $\beta = 3.350918872$. As we shall see,
it is more convenient to express results in terms of $p_0$ and $Q_0$
because our main aim is not to explicitly obtain an expansion for
these quantities in powers of $1/\sigma$, but rather to determine
whether or not the singularity in the EOS state coincides
with the singularity in the susceptibility.

Now we see how the critical coupling constants $p_c(d), Q_c(d)$ for
hypercubic lattices are obtained after $\Delta \Phi$ has been
evaluated.   We write
\begin{eqnarray}
{Q \over p} = \Phi_0(Q) + \Delta \Phi (Q,p) \ , \ \ \ \
{1 \over p} = {d\Phi_0 \over dQ} + {d\Delta \Phi(Q,p) \over dQ} 
\end{eqnarray}
and set
\begin{eqnarray}
p=p_0 + \Delta p \ , \hspace{1 in}
Q &=& Q_0 + \Delta Q \ .
\label{PANDQ} \end{eqnarray}
In this analysis we note that $\Delta \Phi \sim \sigma^{-2}$, so that
to obtain results to order $\sigma^{-3}$ we need only consider terms
linear in $\Delta \Phi$, $\Delta p$, or $\Delta Q$.  Then we have
\begin{eqnarray}
{Q_0 \over p_0} + {\Delta Q \over p_0} - {Q_0 \Delta p \over p_0^2}
= \Phi_0(Q_0) + \Delta Q {d\Phi_0(Q_0) \over dQ} + \Delta \Phi (Q_0,p_0) \ ,
\end{eqnarray}
which gives
\begin{eqnarray}
\Delta p = - {p_0^2 \over Q_0} \Delta \Phi (Q_0,p_0) \ .
\label{DELTAP} \end{eqnarray}
Also
\begin{eqnarray}
{1 \over p_0} - {\Delta p \over p_0^2} &=& {d \Phi_0(Q_0) \over dQ}   
+ \Delta Q {d^2 \Phi_0 \over dQ^2} + {d \Delta \Phi (Q_0,p_0) \over dQ} \ ,
\end{eqnarray}
which gives
\begin{eqnarray}
\Delta Q &=& \left[ {d^2 \Phi_0 \over dQ^2} \right]^{-1} \Biggl[
- {\Delta p \over p_0^2} - {d \Delta \Phi \over dQ} \Biggr] \nonumber \\
&=& \left[ {d^2 \Phi_0 \over dQ^2} \right]^{-1} \Biggl[
{\Delta \Phi (Q_0,p_0) \over Q_0} - {d \Delta \Phi \over dQ} \Biggr] \ .
\label{DELTAQ} \end{eqnarray}

\vspace{0.3 in}
\begin{figure}
\centerline{\psfig{figure=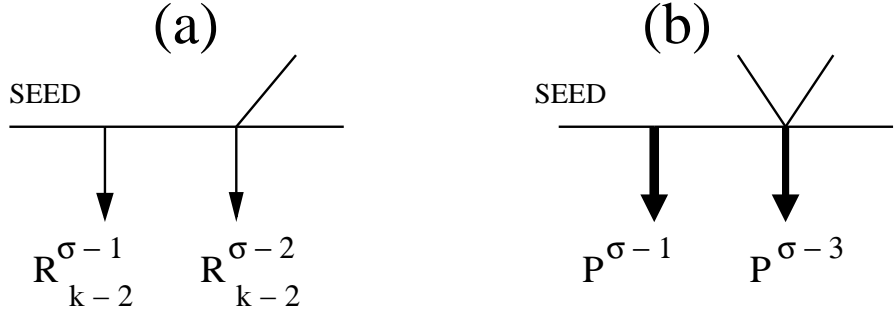}}
\vspace{0.3 in} \caption{Definition of $R^{\sigma-m}_k$ (in panel a) 
and $P^{\sigma-r}$ (in panel b).  Each site has $\sigma+1$ bonds.}
\label{RFIG} \end{figure}

We introduce the following Bethe lattice quantities.  First, 
$R_r^{\sigma-m}$ is defined to be the probability that out of
$\sigma-m$ available bonds, exactly $r$ are $k-1$ connected outward 
(i. e. away from the origin or seed of the cluster) to 
infinity and we set $R_r^{\sigma-1} \equiv R_r$.  (It is also convenient
to set $R_{k-2} \equiv R$).  This definition of $R$ is illustrated in
Fig. \ref{RFIG}.  Similarly we define
$P^{\sigma-m}$ to be the probability that a site is {\em at least}
$k-1$ connected to infinity through the set of $\sigma-m$
of bonds and $P^\sigma \equiv P$.
Since $Q$ is of order $1/\sigma$, we see that $P^{\sigma-m}$
is of order $1/\sigma^0$ (assuming $m$ is of order unity).  One
can likewise show that under this same assumption $R_r^{\sigma-m}$
is also of order unity. Note when the superscript on $P$ is
of the form $\sigma-n$ then the superscript indicates the number
of bonds, as defined above.  When the superscript is is a purely numerical 
value like "2", then 
this indicates an exponent: $P^2=P \times P$.

We now express all the above Bethe lattice quantities in terms of
the canonical variables, $R$, $P$, and $Q_0\equiv p_0P$. We have

\begin{eqnarray}
P^{\sigma-r} &=& 1 - \sum_{m=0}^{m=k-2} {(\sigma-r)! \over
(\sigma-r-m)! m!} Q_0^m (1-Q_0)^{\sigma-r-m} \ ,
\end{eqnarray}
from which we obtain, correct to first order in $1/\sigma$ that
\begin{eqnarray}
P^{\sigma-r} &=& P -rQ_0 R + {\cal O} (\sigma^{-2}) \ .
\end{eqnarray}
Also, 
\begin{eqnarray}
R_m^{\sigma-1-r} &=& Q_0^m {(\sigma-1-r)! \over m! (\sigma -1 -r -m)!}
(1-Q_0)^{\sigma -1 - r -m}
\label {RMEQ}
\\ &=&
R_m \Biggl[ 1 + r Q_0 -{rm \over \sigma} \Biggr] + {\cal O}(\sigma^{-2}) \ ,
\end{eqnarray}
\begin{eqnarray}
{dP \over dQ} &=& \sigma R \ ,
\end{eqnarray}
and
\begin{eqnarray}
{dR_m \over dQ} &=& {(\sigma-1)! \over (m-1)! (\sigma-m-1)!}
Q_0^{m-1} (1-Q_0)^{\sigma-m-1} - {(\sigma-1)! \over m! (\sigma-m-2)!}
Q_0^m (1-Q_0)^{\sigma-m-2} \nonumber \\ &=& {1 \over 1-Q_0} \Biggl[
{(\sigma-1)! \over (m-1)! (\sigma-m-1)!}
Q_0^{m-1} (1-Q_0)^{\sigma-m} - {(\sigma-1)! \over m! (\sigma-m-2)!}
Q_0^m (1-Q_0)^{\sigma-m-1} \Biggr] \nonumber \\ &=& {1 \over 1-Q_0} \Biggl[
(\sigma -m) R_{m-1} - (\sigma -m -1) R_m \Biggr] \nonumber \\ 
&=& (\sigma -m +Q)(R_{m-1}-R_m) +R_m + {\cal O}(1/\sigma ) \ .
\end{eqnarray}

\section{$1/\sigma$ EXPANSION FOR THE EQUATION OF STATE}

\begin{figure}
\centerline{\psfig{figure=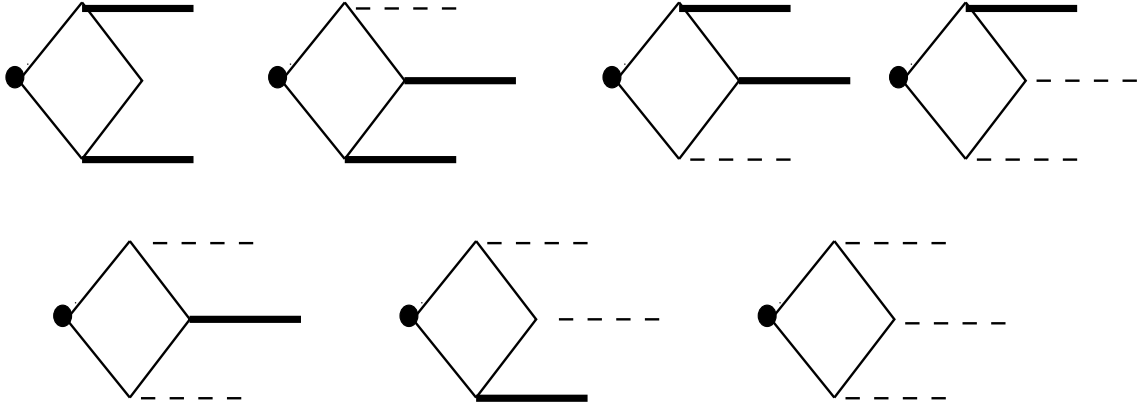}}
\vspace{0.3 in} \caption{Configurations of diagram (a) of Fig.
\protect{\ref{F13}}
which have two paths to infinity.  The heavy line is a bond
which is $k-1$-connected to infinity and carries a factor $P^{\sigma -1}$.
The dashed line carries a factor $R_{k-2}$ which may survive culling if
it is connected to two live bonds.}
\label{RFIG1} \end{figure}

\begin{figure}
\centerline{\psfig{figure=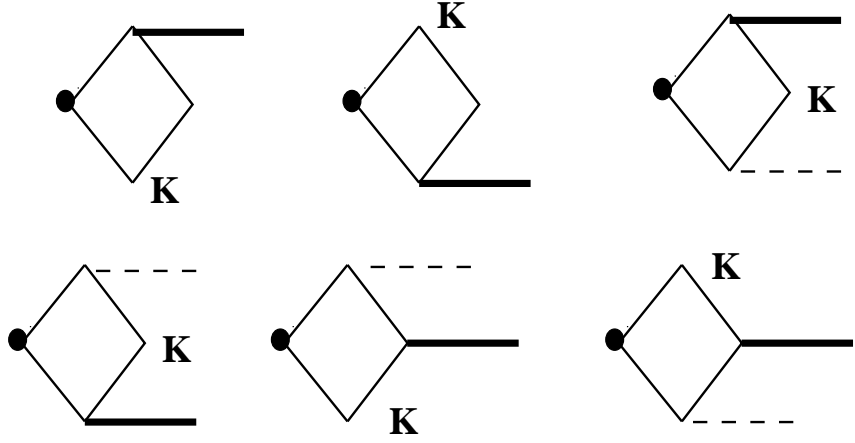}}
\vspace{0.3 in} \caption{As in Fig. \protect{\ref{RFIG1}},
configurations of diagram (a) of Fig. \protect{\ref{F13}} 
which have one path to infinity.
The symbol K denotes a bond which is definitely culled
and therefore carries a factor $(1-P^{\sigma-1}-R_{k-2})$.}
\label{RFIG2} \end{figure}

\begin{figure}
\centerline{\psfig{figure=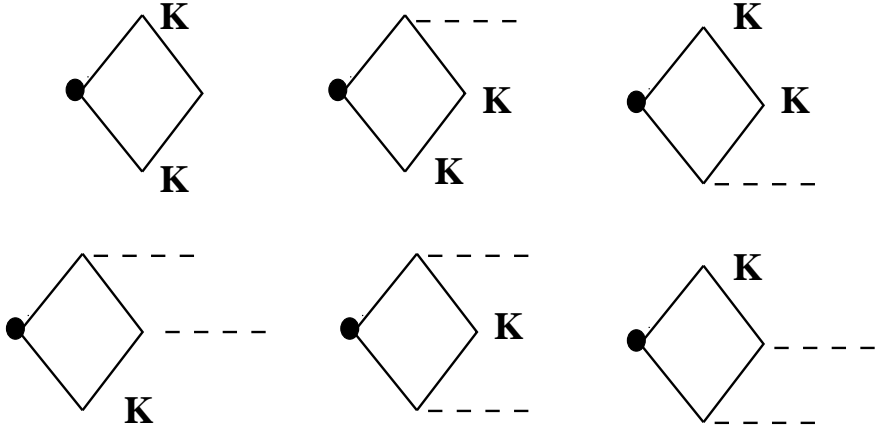}}
\vspace{0.3 in} \caption{As in Figs. \protect{\ref{RFIG1}} and
\protect{\ref{RFIG2}}, configurations of diagram (a) of Fig.
\protect{\ref{F13}} which have no paths to infinity.}
\label{RFIG3} \end{figure}

In this section we will implement the $1/\sigma$ expansion for the
EOS.  Note that we are considering the effect of having a square
(or hexagon) of bonds with one vertex at the seed site. (Fig. \ref{SQ7}. )
All structures emanating from this square (or hexagon) or from
the other $\sigma-2$ bonds which intersect the seed site,
may be assumed to be tree-like because the occurrence of more than
one loop only influences terms of relative order $1/\sigma^4$
and we do not consider this order.  Accordingly we implement
Eq. (\ref{SQPHI}).  For squares we have
$f(n) = [(\sigma-1)^2 p^4 /2]\delta f(n)$, where
$n$ indicates the diagram as labeled in Eq. (\ref{SQPHI}).  Since the 
insertion of the square (or hexagon) can contribute a maximum of two 
more paths to the boundary (or infinity) at the seed site, we will 
break up diagram (a) of Fig. 3 into factors associated with having two, one, and zero
paths to infinity as illustrated in Figs. \ref{RFIG1}, \ref{RFIG2},
and \ref{RFIG3}, respectively. Thereby we find

\vspace{0.3 in}

\vspace{0.3 in}

\vspace{0.3 in}

\begin{eqnarray}
\delta f(a) &=& \left( R_{k-3}^{\sigma-2} + R_{k-2}^{\sigma-2} 
+ P^{\sigma-2} \right) \left[ (P^{\sigma-1})^2
+ 2 R (P^{\sigma-1})^2 + 3 R^2 P^{\sigma-1} + R^3 \right] \nonumber \\ &+&
[R_{k-2}^{\sigma-2} + P^{\sigma-2}] [2 + 4 R] P^{\sigma-1}
[1 - P^{\sigma-1} - R] \nonumber \\ &+&
P^{\sigma-2} \Biggl[ (1- P^{\sigma-1} - R)^2 + 2 R (1- P^{\sigma-1} - R)^2
+ 3 R^2(1- P^{\sigma-1} - R) \Biggr]
\nonumber \\ &=&
R_{k-3}^{\sigma-2} \Biggl[ (P^{\sigma-1})^2 + 2 R
(P^{\sigma-1})^2 + 3 R^2 P^{\sigma-1} +
R^3 \Biggr] + R_{k-2}^{\sigma-2} +P^{\sigma-2} 
\nonumber \\ &&
- R_{k-2}^{\sigma-2} \Biggl[ (1- P^{\sigma-1} - R)^2
+2 R (1- P^{\sigma-1} - R)^2 \nonumber \\ &&
+ 3 R^2 (1- P^{\sigma-1} - R ) \Biggr]
\end{eqnarray}
In constructing this expression we noted that if the square
had two paths to infinity, then the remaining $\sigma -2$ bonds emanating
from the seed site had to have at least $k-3$ paths to infinity
which we take into account with the factor of the form
$R_{k-3} + R_{k-2} + P$.  ($P$ is the
probability of having at least $k-1$ paths to infinity.)
The factor $R_{k-2}+P$ and $P$ take account of
having at least $k-2$ or $k-1$ paths to infinity, respectively.
Similarly, for the other diagrams we have
\begin{eqnarray}
\delta f(b) &=& [P^{\sigma-2} + R_{k-2}^{\sigma-2}]
P^{\sigma-1} (1+R + R^2) 
+ P^{\sigma-2} [ 1 - P^{\sigma-1} (1+R + R^2)]
\nonumber \\ &=& P^{\sigma-2} + R_{k-2}^{\sigma-2} P^{\sigma-1}
(1+R + R^2)
\end{eqnarray}
\begin{eqnarray}
\delta f(d) &=& [P^{\sigma-2} + R_{k-2}^{\sigma-2}
+R_{k-3}^{\sigma-2}] (P^{\sigma-1} )^2 (1+R) \nonumber \\ &&
+ [P^{\sigma-2} + R_{k-2}^{\sigma-2}] \Biggl[ P^{\sigma-1}
[1-P^{\sigma-1}(1+R)] + (1-P^{\sigma-1}) 
P^{\sigma-1} (1+R)\Biggr] \nonumber \\ &&
+ P^{\sigma-2} [1-P^{\sigma-1}] [1 - P^{\sigma-1}
(1+R)] \nonumber \\ &=&
P^{\sigma-2} + R_{k-2}^{\sigma-2} + R_{k-3}^{\sigma-2}
(P^{\sigma-1} )^2 (1+R) \nonumber \\ &&
- R_{k-2}^{\sigma-2}
[1-P^{\sigma-1}] [1 - P^{\sigma-1}(1+R)]
\end{eqnarray}
\begin{eqnarray}
\delta f(i) &=& P^{\sigma-2} [1-P][1-P^{\sigma-1}_\infty
(1+R) - P R^2]\nonumber \\ &&
+ (R_{k-2}^{\sigma-2} + P^{\sigma-2}) \Biggl[ [1-P]
[P^{\sigma-1}(1+R) + P R^2] \nonumber \\ &&
+ P [ 1 - P^{\sigma-1}_\infty (1+R) - P
R^2)]\Biggr] \nonumber \\ && 
+ (R_{k-3}^{\sigma-2}  + R_{k-2}^{\sigma-2} + P^{\sigma-2})
P [P^{\sigma-1}(1+R)+P R^2]\Biggr]
\nonumber \\ &=& R_{k-2}^{\sigma-2} + P^{\sigma-2} +
R_{k-3}^{\sigma-2} P [P^{\sigma-1}(1+R)+P
R^2] \nonumber \\ && - R_{k-2}^{\sigma-2} (1-P)
\Biggl[1 - P^{\sigma-1} (1+R)-P R^2 \Biggr]
\end{eqnarray}
\begin{eqnarray}
\delta f(j) &=& [P^{\sigma-2} + R_{k-2}^{\sigma-2}]
[P^{\sigma-1} (1+R) + P R^2]
+ P^{\sigma-2} [1- P^{\sigma-1} (1+R)
- P R^2] \nonumber \\ &=&
P^{\sigma-2} + R_{k-2}^{\sigma-2} [P^{\sigma-1} 
(1+R) + P R^2]
\end{eqnarray}
\begin{eqnarray}
\delta f(k) &=& [P^{\sigma-2} + R_{k-2}^{\sigma-2}
+R_{k-3}^{\sigma-2}] P P^{\sigma-1}(1+R) \nonumber \\ &&
+ [P^{\sigma-2} + R_{k-2}^{\sigma-2}] \Biggl[ P^{\sigma-1}
[1-P](1+R) + P
[1-P^{\sigma-1} (1+R)\Biggr] \nonumber \\ &&
+ P^{\sigma-2} [1-P] [1 - P^{\sigma-1}
(1+R)] \nonumber \\ &=&
P^{\sigma-2} + R_{k-2}^{\sigma-2} + R_{k-3}^{\sigma-2}
P P^{\sigma-1} (1+R) \nonumber \\ &&  - R_{k-2}^{\sigma-2}
[1-P] [1 - P^{\sigma-1}(1+R)] \ .
\end{eqnarray}
\begin{eqnarray}
\delta f(m) &=& P^{\sigma-2}[1-P^{\sigma-1} -
RP]^2 \nonumber \\ && +
2 [P^{\sigma-2} + R_{k-2}^{\sigma-2}][1-P^{\sigma-1} -
RP][P^{\sigma-1} + R P]
\nonumber \\ &&
+ [R_{k-3}^{\sigma-2} + R_{k-2}^{\sigma-2} + P^{\sigma-2}]
[P^{\sigma-1} + R P]^2 \nonumber \\ &=&
P^{\sigma-2} + R_{k-2}^{\sigma-2} + R_{k-3}^{\sigma-2}
[P^{\sigma-1} + R P]^2 - R_{k-2}^{\sigma-2}
[1-P^{\sigma-1} - RP]^2 
\end{eqnarray}
\begin{eqnarray}
\delta f(n) &=&
P^{\sigma-2} + R_{k-2}^{\sigma-2} + R_{k-3}^{\sigma-2}
P^{\sigma-1} [P^{\sigma-1} + RP]
\nonumber \\ &&
- R_{k-2}^{\sigma-2} [1-P^{\sigma-1}] 
[1 - P^{\sigma-1}-P R]
\end{eqnarray}
\begin{eqnarray}
\delta f(q)&=& P^{\sigma-1} \Biggl[ 1 - P^{\sigma-1}
(1+R+R^2) - R^3P \Biggr] \nonumber \\ &&
+ (R+P^{\sigma-1})\Biggl[ P^{\sigma-1}(1+R)
+R^2) + R^3 P \Biggr] \nonumber \\ &=&
P^{\sigma-1} + R \Biggl[ P^{\sigma-1}(1+R
+R^2) + R^3 P \Biggr]
\end{eqnarray}
\begin{eqnarray}
\delta f(r) &=& [P^{\sigma-1} + R] P^{\sigma-1} 
[1+R+R^2]
+ P^{\sigma-1} [1- P^{\sigma-1} (1+R+R^2)]
\nonumber \\ &=& P^{\sigma-1} + R P^{\sigma-1} 
[1+R+R^2]\,.
\end{eqnarray}
With these results we are now ready to evaluate Eq. (\ref{SQPHI}) and 
it reads 
\begin{eqnarray}
\Phi_{\rm Sq} &=& {(\sigma-1)^2 p^4 \over 2} \Biggl\{ 
R_{k-3}^{\sigma-2} [ R^3 + 3R^2 P^{\sigma-1}
-3R^2 P^2] -2R^4 P \nonumber \\ && \
+ R^{\sigma-2}_{k-2} [ R^3-3R^2P^{\sigma-1} +3R^2 P^2 ] \Biggr\}  \ .
\label{PHISQEQ} \end{eqnarray}

Since the contributions from hexagons only need to be evaluated to leading
order in $1/\sigma$, the configuration of vacant bonds is irrelevant
and we may omit the superscripts on $R$ and $P$.  Accordingly, for hexagons
it is convenient to classify (as indicated by the subscript) the value of
$m$, the number of paths to infinity.  For diagram $x$ we write
\begin{eqnarray}
f_m(x) =2 \sigma^3 p^6 \delta f_m (x) \ ,
\end{eqnarray}
where $2 \sigma^3$ is the number of hexagons (to lowest order in
$1/\sigma$) that can be attached to a bond and
\begin{eqnarray}
\delta f_2(a) &=& P^2 [1 + 2R + 3 R^2 
+ 4 R^3] + 5 R^4 P + R^5
\end{eqnarray}
\begin{eqnarray}
\delta f_1(a) &=& 2P [ 1 - P - R][1+2 R
+3 R^2 + 4 R^3]
\end{eqnarray}
\begin{eqnarray}
\delta f_0(a) &=& [1-P - R]^2 [1 + 2 R 
+ 3 R^2 + 4 R^3] + 5 R^4 [1-P-R]
\end{eqnarray}
\begin{eqnarray}
\delta f_1(b) &=& P[1 +R + R^2 + R^3
+ R^4 + R^5]
\end{eqnarray}
\begin{eqnarray}
\delta f_0(b) &=& [1-P-R]
[1 +R + R^2 + R^3 + R^4]
+ R^5[1-P]
\end{eqnarray}
\begin{eqnarray}
\delta f_2(c) &=& P^2
[1 +R + R^2 + R^3 + R^4]
\end{eqnarray}
\begin{eqnarray}
\delta f_1(c) &=& [1-P]P
[1 +R + R^2 + R^3 + R^4] \nonumber \\ &&
+ P[1-P -R]
[1 +R + R^2 + R^3 ] + P[1-P]R^4
\end{eqnarray}
\begin{eqnarray}
\delta f_0(c) &=& [1-P][1-P-R]
[1 +R + R^2 + R^3 ] + [1-P]^2 R^4
\end{eqnarray}
\begin{eqnarray}
\delta f_2(d) &=& P^2 [1 +R + R^2 + R^3 ]
[1 +R ]
\end{eqnarray}
\begin{eqnarray}
\delta f_1(d) &=& P[1-P -RP]
[1 +R + R^2 + R^3 ] \nonumber \\ &&
+ P [1+R][(1-P - R) (1 +R + R^2)
+ (1-P)R^3]
\end{eqnarray}
\begin{eqnarray}
\delta f_0(d) &=& [1-P - RP][
(1-P - R) (1 +R + R^2) + (1-P)R^3 ]
\end{eqnarray}
\begin{eqnarray}
\delta f_2(e) &=& [P (1 +R + R^2)]^2
\end{eqnarray}
\begin{eqnarray}
\delta f_1(e) &=&
2P [1 +R + R^2] [(1-P -R)(1+R)
+ R^2 (1-P)]
\end{eqnarray}
\begin{eqnarray}
\delta f_0(e) &=& [(1-P-R)(1+R) + R^2
(1-P)]^2
\end{eqnarray}
\begin{eqnarray}
\delta f_1(f) &=& P [1 +R + R^2 +R^3 + R^4]
\end{eqnarray}
\begin{eqnarray}
\delta f_0(f) &=&[(1-P-R) (1 +R + R^2 +R^3) + (1-P)R^4]
\end{eqnarray}
\begin{eqnarray}
\delta f_2(g) &=& P^2 [1 +R + R^2 +R^3]
\end{eqnarray}
\begin{eqnarray}
\delta f_1(g) &=& P(1-P) [1 +R + R^2 +R^3]
\nonumber \\ &&
+P[(1-P-R) (1 +R + R^2) + (1-P) R^3]
\end{eqnarray}
\begin{eqnarray}
\delta f_0(g) &=& (1-P)[
(1 +R + R^2 )(1-P -R) + (1-P)R^3 ]
\end{eqnarray}
\begin{eqnarray}
\delta f_2(h) &=& P^2 (1 +R + R^2 )
(1 +R)
\end{eqnarray}
\begin{eqnarray}
\delta f_1(h) &=& (1-P-RP)P
(1 +R + R^2 )
\nonumber \\ &&
+ P (1 +R)(1-P
- RP- R^2P)
\end{eqnarray}
\begin{eqnarray}
\delta f_0(h) &=& (1-P-RP)
(1-P - RP- R^2P) \ ,
\end{eqnarray}
where the numbering of contributions is as in Fig. \ref{H1a}.
Thus for the contribution of hexagons to the EOS (indicated by the superscript "H"), the sum over diagrams gives
\begin{eqnarray}
\Phi^H_2 &=& f_2(a) -2 f_2(c) -2 f_2(d) - f_2(e) + 2 f_2(g)
+ 2 f_2(h) \nonumber \\ &=& 2 \sigma^3 p^6 \Biggl(
-5R^4 P_\infty^2 + 5 R^4 P_\infty + R^5\Biggr) \ . 
\end{eqnarray}
\begin{eqnarray}
\Phi^H_1 &=& f_1(a) -2 f_1(b) -2 f_1(c) - 2 f_1(d) - f_1(e)
+2 f_1(f) +2 f_1(g) + 2 f_1(h) \nonumber \\ &=& 2 \sigma^3 p^6 \Biggl(
10R^4 P_\infty^2 - (10 R^4 +2 R^5) P_\infty\Biggr) \ ,
\end{eqnarray}
and
\begin{eqnarray}
\Phi^H_0 &=& f_0(a) -2 f_0(b) -2 f_0(c) - 2 f_0(d) - f_0(e)
+2 f_0(f) +2 f_0(g) + 2 f_0(h) \nonumber \\ &=& 2 \sigma^3 p^6 \Biggl(
-5R^4P_\infty^2 + (5R^4 + 2R^5)P_\infty -R^5\Biggr) \ .
\end{eqnarray}
It is a check on our results that $\sum_n \Phi_n^H = 0$.
Now we match each of these contributions to possible configurations of
the other $\sigma-2$ bonds from the seed site.  Thus, from hexagons we get
\begin{eqnarray}
\Phi_{\rm Hex} &=& P \Phi^H_0 + (P + R) \Phi^H_1
+ (P+ R+R_{k-3}) \Phi^H_2 \nonumber \\
&=& R_{k-3} \Phi_2^H - R \Phi_0^H \ . 
\label{PHIHEX} \end{eqnarray}

\section{$1/\sigma$ EXPANSION FOR THE SUSCEPTIBILITY}

\subsection{Formulation of the Susceptibility}

In this section we consider the $1/\sigma$ expansion for the
two-point ($i,j$) susceptibility, $\chi_{ij}$. For $k$-core
percolation we calculate this
quantity in the ordered phase, because for the Bethe lattice there
are no finite $k$-core clusters.  Here we define
$\chi=\sum_j \chi_{ij}$, where
\begin{eqnarray}
\chi_{ij} = \langle \nu_i \nu_j \rangle - \langle \nu_i \rangle
\langle \nu_j \rangle \ ,
\label{CHIEQ} \end{eqnarray}
where $\nu_i$ is an indicator variable which is unity if the
site $i$ is in the $k$-core and is zero otherwise.
Also the angle brackets indicate an average of configurations of bonds
in which each bond is independently present with probability $p$
and absent with probability $1-p$.

\begin{figure}
\centerline{\psfig{figure=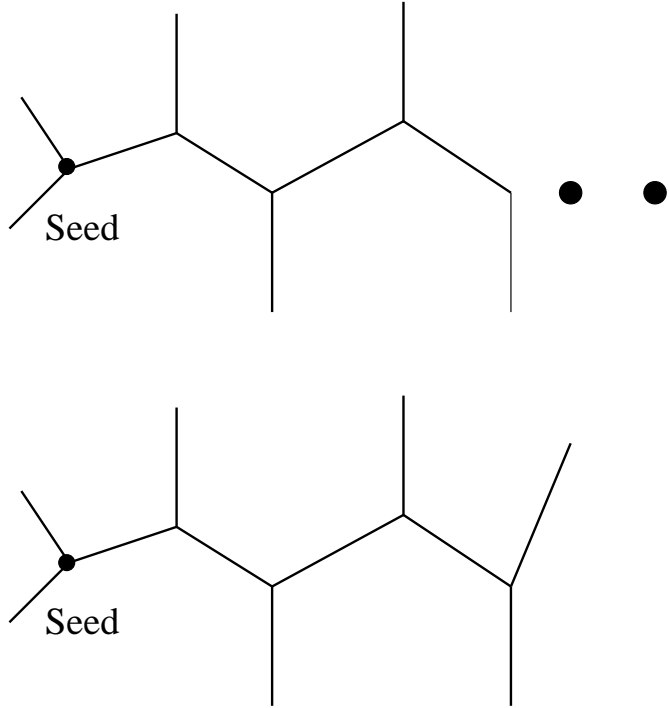}}
\vspace{0.3 in} \noindent
\caption{Configurations for $k=3$ illustrating "truncation" of
diagrams.  Here the side branches are assumed to be
$k-1$-connected to infinity.
In the top panel a configuration is shown where the
$k$-core depends on the state of the further bonds indicated by
filled circles.  In the lower panel, the configuration is $k=3$
connected no matter what happens further down the chain.
Since the $k$-core does not depend on further bond occupation
probabilities  this diagram can not be extended.}
\label{sprout} \end{figure}

Now we formulate the $1/\sigma$ expansion for the ordered phase of
$k$-core percolation.  On the Bethe lattice the long-range
part of this correlation function comes from the probability of
configurations in which two sites are only in the $k$-core by virtue of
the presence of a path of occupied bonds connecting sites $i$ and $j$.
(Of course each site on this path must belong to the $k$-core.)
In this connection, it is important that the sites not be in the
$k$-core if any bond is removed.  Such contributions are either already
counted in lower order or are canceled when the second term in
Eq. (\ref{CHIEQ}) is subtracted off.  Thus the configuration in panel (a)
of Fig. \ref{sprout} is an allowed configuration contributing to
$\chi_{ij}$, but that in panel (b) can not be extended. In Ref.
\onlinecite{SLC},
this consideration led to equivalently restricting the sum to the "corona."
Thus, as the path is progressively lengthened, if we reach a point where
the origin of the path is certainly in the $k$-core no matter how the
path is extended, then this path is said to be "truncated," and is
discarded as not contributing to the singularity in $\chi$.

\vspace{0.3 in}

For the present work we obtain the results of Ref. \onlinecite{SLC}
as follows.  For a path to satisfy the "corona" constraint, it must
consist of a path for which each vertex is $k-2$ connected to infinity
if the bonds along the path are not considered.  With this construction
one sees that at each vertex, whether the whole structure is or is not
$k$-connected to infinity depends on what happens further down the
path.  Thus, for the Bethe lattice each vertex in the path leading
from site $i$ to site $j$ carries a factor $pR_{k-2}^{\sigma-1}=
pR_{k-2}=pR$.  When $\chi_{ij}$ is summed over $j$, one obtains a
geometric series in the ratio, $r$, given by
\begin{eqnarray}
r &=& \sigma p R_{k-2} \ ,
\end{eqnarray}
which coincides with the singularity in the EOS~\cite{SLC}. 

\begin{figure}
\centerline{\psfig{figure=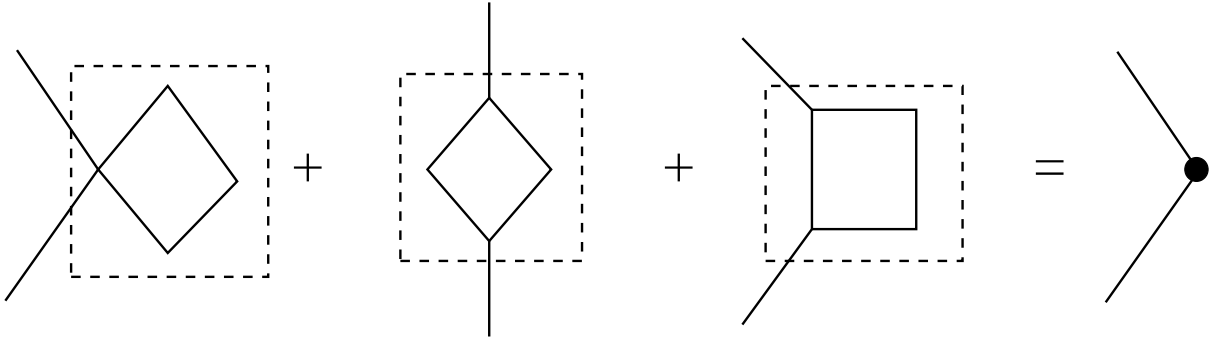}}
\vspace{0.3 in} \caption{The factors associated with the dashed
box in the first three diagrams renormalize the vertex (the filled
circle) in the last diagram. We label these contributions
SQ1, SQ2, and SQ3, respectively.} 
\label{VERTEX} \end{figure}

A convenient way to describe the $1/\sigma$ expansion as applied
to the susceptibility, is to regard the $1/\sigma$ corrections 
as a vertex correction, as illustrated in  Fig. \ref{VERTEX}
for insertions of 
\noindent
a square.  This vertex correction amounts to replacing $R_{k-2}$ by
$R_{k-2} + \Delta R$, where $\Delta R$ is
the contribution from the dashed box (which replaces each vertex).
Analogous corrections arise at relative order
$1/\sigma^3$ from insertions of hexagons. However,
insertions involving two or more loops lead to corrections
of order $1/\sigma^4$ and higher and are beyond the scope of the
present calculation.  
(Note that for the last diagram, SQ3, we need an extra factor of 2
because the two sides of the square are inequivalent, so that it
occurs $(\sigma -1)^2$ ways.)  The diagrams for a square insertion into
a chain diagram for the susceptibility are shown more explicitly below.

The condition for a divergent susceptibility is that
the perturbed ratio $r$ be unity, or
\begin{eqnarray}
1 &=& r = \sigma p R + \sigma p \Delta R \ .
\end{eqnarray}
We now evaluate the ratio $r$ at the singularity in the EOS,
so that $p$ and $Q$ are given by Eqs. (\ref{PANDQ}), (\ref{DELTAP}),
and (\ref{DELTAQ}).  Working to leading order we write
\begin{eqnarray}
r &=& \sigma (p_0 + \Delta p) \left[ R(Q_0) + {dR \over dQ_0} \Delta Q
\right] + \sigma p \Delta R \nonumber \\
&=& 1 + {\Delta p \over p_0} + {1 \over R} {dR \over dQ_0} \Delta Q + \sigma p_0
\Delta R \nonumber \\ &=&
1 - {p_0 \over Q_0} \Delta \Phi + {1 \over R} {dR \over dQ_0} \left[
\sigma { dR \over dQ_0} \right]^{-1} \left[ {\Delta \Phi \over Q_0} - {d \Delta
\Phi \over dQ} \right] + \sigma p_ \Delta R \nonumber \\ &=&
1 + \left[ \Delta R - {d \Delta \Phi \over \sigma dQ} \right] R^{-1} \ .
\label{ratioeq} \end{eqnarray}
Here we noted that for the Bethe lattice $\sigma p_0 R(Q_0)=1$
and $d\Phi /dQ = \sigma R$. 
There are now three scenarios, depending on whether the ratio $r$
is a) less than, b) equal to, or c) greater than unity.
In case a) the discontinuity in the EOS preempts the
divergence of the susceptibility and the transition is a conventional
first order transition, except for a fractional power law in the EOS
for $p$ above threshold.  Case b) (consistent with
the expansion up to order $1/\sigma^3$) indicates
that the coincidence of the singularities in
the EOS and the susceptibility is robust and survives for large
but finite spatial dimension. 

\subsection{Diagrams for the Susceptibility}

Now we consider the effect of inserting squares or hexagons
into the Bethe lattice diagrams for the susceptibility.
\vspace{0.3 in}
\begin{figure}
\centerline{\psfig{figure=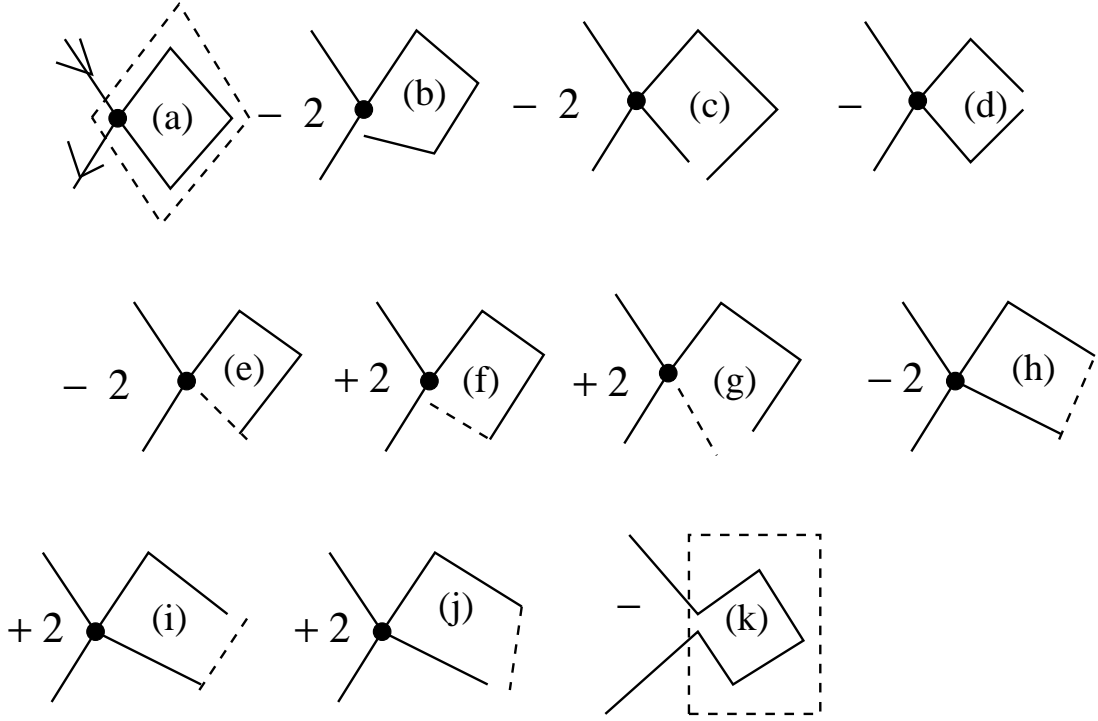}}
\vspace{0.3 in} \caption{Diagrams for Square \#1 on either H or B
lattices.  Diagram (k) occurs on the Bethe lattice
but must be subtracted off because its counterpart on
the hypercubic lattice has a different topology.}
\label{D7} \end{figure}
\begin{figure}
\centerline{\psfig{figure=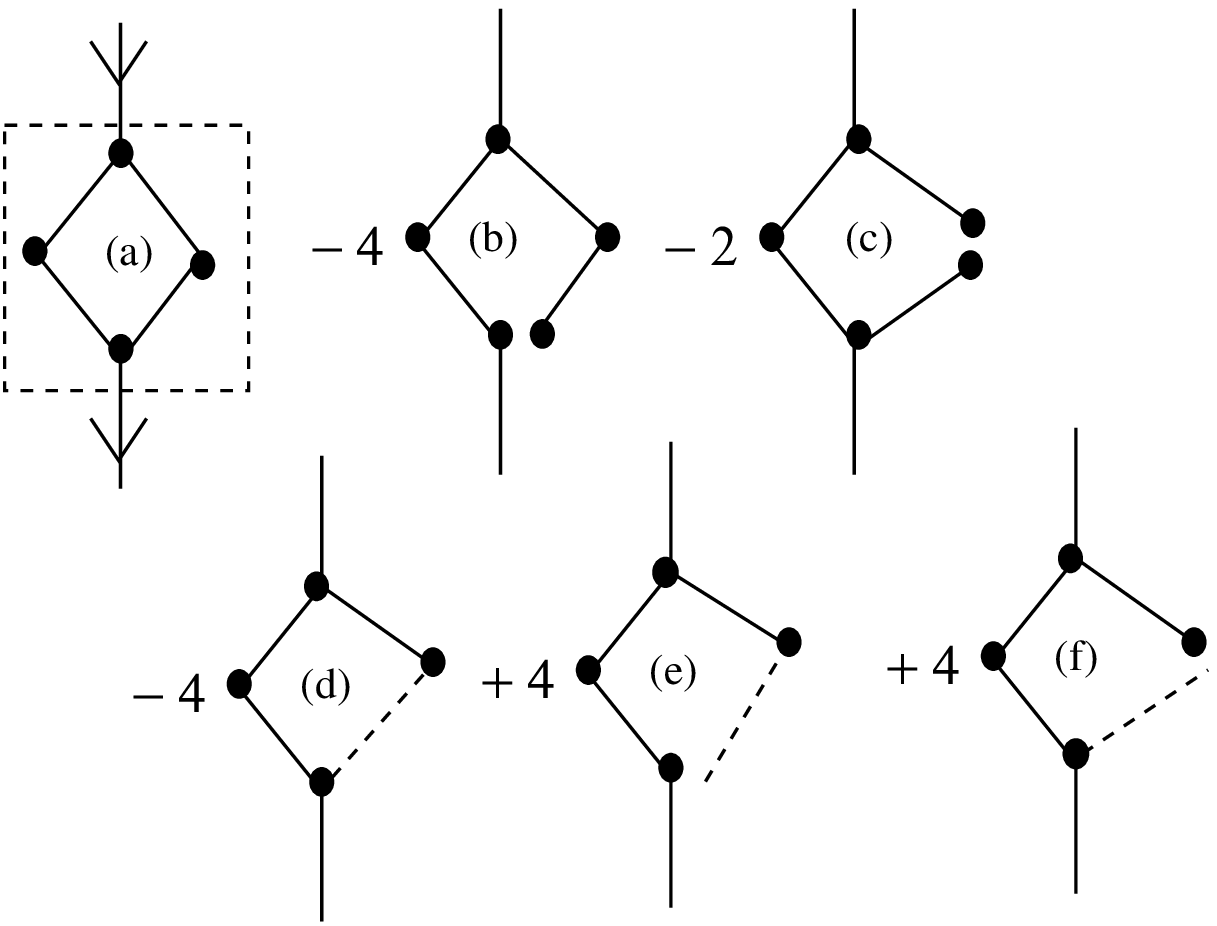}}
\vspace{0.3 in} \caption{Diagrams for Square \#2 on
either H or B lattices.}
\label{D8} \end{figure}
\vspace{0.3 in}
\begin{figure}
\centerline{\psfig{figure=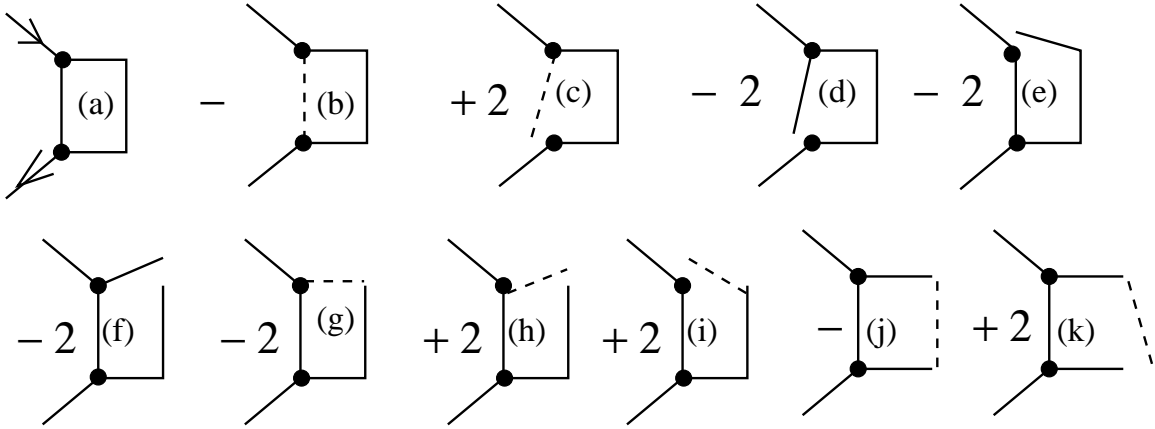}}
\vspace{0.3 in} \caption{Diagrams for Square \#3 on either H or B lattices.}
\label{D9} \end{figure}
For squares we write
\begin{eqnarray}
\Delta r_{\rm Sq} &=& (\sigma-1)^2 \frac{p^4}{2} \Delta R / R_{k-2} \ ,
\end{eqnarray}
where the square insertions are discussed below.  Similarly, for  
hexagons we write
\begin{eqnarray}
\Delta r_{\rm Hex} &=& 2p^6 \sigma^3 \Delta R / R_{k-2} \ ,
\end{eqnarray}
where the hexagon insertions are also discussed below.

\subsection{Percolation}

As an example we first carry out the calculation for the
two-point susceptibility of percolation and we confine our
attention to the disordered phase, i.e. $P_{\infty}=0$.  
(The calculation for the
ordered phase is quite similar.)  First, consider the squares.
The sum of the contributions from diagrams (a) through (j)
of Fig. \ref{D7} vanishes.  
For diagram (k) there are $N_{\rm Sq}=(\sigma-1)^2/2$ ways to
attach the pseudosquare (which can be traversed in two senses).
So the dashed box of diagram (k)
carries the factor
\begin{eqnarray}
\Delta r^{\#1}  &=& - p^4 \sigma^{-2} [ 1 - (2/\sigma)] \ .
\end{eqnarray}
The contribution from Fig. \ref{D8} is $\Delta r^{\#2}= - (\sigma-1)^3 / \sigma^5$
and that from Fig. \ref{D9} is $\Delta r^{\#3}= -(1/2)(\sigma-1)^3 /\sigma^5$. 
So the total contribution from squares is
\begin{eqnarray}
\Delta r_{\rm Sq} &=& - (5/2) \sigma^{-2} + (13/2) \sigma^{-3} \ ,
\end{eqnarray}
The contribution from hexagons is
\begin{eqnarray}
\Delta r_{\rm Hex} &=& - 14 \sigma^{-3} \ ,
\end{eqnarray}
When these contributions are summed, one finds that the
threshold for the divergence of the susceptibility agrees
to order $1/\sigma^3$ with that in Eq. (\ref{GRPERC}) found from
the EOS and with the previous work of Gaunt and Ruskin~\cite{HDPERC}. 
Of course, this was hardly surprising, but the present
calculation illustrates the  more complicated calculation
needed for $k$-core percolation.

\subsection{$k$-Core Percolation, Squares}

We now consider the contributions to the susceptibility from
insertions of squares as in Fig. \ref{D7}. We calculate
the factor $\Delta R$ associated with the dashed box.
In the results given below we include the prefactors (e. g. $\pm 2$
or $\pm 1$) written in Fig. \ref{D7}.  We find
\begin{eqnarray}
\Delta R (a) &=& {x(\sigma-2)\over 2 \sigma }
\Biggl\{ R_{k-4}^{\sigma-3} \Biggl[ (P^{\sigma-1})^2
+ 2 R (P^{\sigma-1})^2 +3 R^2 P^{\sigma-1} + 
R^3 \Biggr] \nonumber \\ &&
+ R_{k-3}^{\sigma-3} \Biggl[ 2P^{\sigma-1} (1-P^{\sigma-1} 
- R) + 4 RP^{\sigma-1}
(1-P^{\sigma-1}-R) \Biggr] \nonumber \\ &&
+ R_{k-2}^{\sigma-3} \Biggl[ (1-P^{\sigma-1}
-R)^2 (1+ 2 R)
+3 R^2 (1-P^{\sigma-1}-R) \Biggr] \Biggr\} \ .
\end{eqnarray}
where $x=(\sigma-1)^2 p^4$
and we used $N_{\rm Sq} = (\sigma-1)^2/2$.  Also,
the factor $(\sigma-2)/\sigma$ takes account of the fact that the
line leaving the square has only $\sigma-2$ choices because it
has to avoid the square. This factor appears for the other diagrams,
except that for (b) and (f) the counting is less obvious and is
discussed in Appendix A.  We have
\begin{eqnarray}
\Delta R(b) &=& - x {(\sigma-2)\over \sigma} \Biggl\{
R_{k-2}^{\sigma-2}[1 - P^{\sigma-1}
(1 + R + R^2) - R^3 P] 
\nonumber \\ && \
+ R_{k-3}^{\sigma-2}[P^{\sigma-1} (1 + R + R^2)
+ R^3P] \Biggr\} \ ,
\end{eqnarray}
\begin{eqnarray}
\Delta R (c) &=& - x {(\sigma-2)\over \sigma } \Biggl\{
R_{k-2}^{\sigma-3} (1-P)
[1-P^{\sigma-1}(1+ R) - R^2 P ] \nonumber \\ &&
+ R_{k-3}^{\sigma-3} P [1-P^{\sigma-1}
(1 + R) - R^2 P ]
\nonumber \\ &&
+ R_{k-3}^{\sigma-3} (1-P) [P^{\sigma-1} (1 + R)
+ R^2 P ] \nonumber \\ &&
+ R_{k-4}^{\sigma-3} P [ P^{\sigma-1}(1 + R)
+ R^2 P ] \Biggr\} \ ,
\end{eqnarray}
\begin{eqnarray}
\Delta R(d) &=& - x {(\sigma-2)\over 2 \sigma } \Biggl\{
R_{k-2}^{\sigma-3} [1-P^{\sigma-1} - RP]^2
\nonumber \\ &&
+2 R_{k-3}^{\sigma-3} [1- P^{\sigma-1} - RP]
[P^{\sigma-1} + R P]
+ R_{k-4}^{\sigma-3} [P^{\sigma-1} + R P]^2 \Biggr\} \ ,
\end{eqnarray}
\begin{eqnarray}
\Delta R (e) &=& - x {(\sigma-2)\over \sigma } \Biggl\{
R_{k-2}^{\sigma-3} [1-P^{\sigma-1}(1 + R + R^2) ] 
+R_{k-3}^{\sigma-3} P^{\sigma-1} [1+R+R^2] \Biggr\}
\end{eqnarray}
\begin{eqnarray}
\Delta R(f) &=& x {(\sigma-2) \over \sigma } \Biggl\{
R_{k-2}^{\sigma-2} [1-P^{\sigma-1}(1 + R + R^2) ]
+ R_{k-3}^{\sigma-2} P^{\sigma-1} [1 + R + R^2] \Biggr\}
\end{eqnarray}
\begin{eqnarray}
\Delta R(g) &=& x {(\sigma-2)\over \sigma } \Biggl\{
R_{k-2}^{\sigma-3} [1-P^{\sigma-1}(1+R) - R^2 P]
+ R_{k-3}^{\sigma-3} [P^{\sigma-1} (1 + R)
+ R^2 P ] \Biggr\}
\end{eqnarray}
\begin{eqnarray}
\Delta R(h) &=& - x {(\sigma-2)\over \sigma } \Biggl\{
R_{k-4}^{\sigma-3} (P^{\sigma-1})^2
[1+R] + R_{k-3}^{\sigma-3} [1-P^{\sigma-1} ]
P^{\sigma-1} [1+R] \nonumber \\ &&
+ R_{k-3}^{\sigma-3} P^{\sigma-1}
[1 - P^{\sigma-1} - RP^{\sigma-1}]
\nonumber \\ && + R_{k-2}^{\sigma-3}
[1-P^{\sigma-1}][1-P^{\sigma-1} - R P^{\sigma-1}] \Biggr\} \ ,
\end{eqnarray}
\begin{eqnarray}
\Delta R(i) &=& x {(\sigma-2)\over \sigma } \Biggl\{
R_{k-4}^{\sigma-3} P^{\sigma-1}
[P^{\sigma-1} + RP]
+ R_{k-3}^{\sigma-3} [1-P^{\sigma-1} ] [P^{\sigma-1} + 
R P ] \nonumber \\ &&
+ R_{k-3}^{\sigma-3} P^{\sigma-1} [1 - P^{\sigma-1}
- RP] + R_{k-2}^{\sigma-3}[1-P^{\sigma-1}]
[1-P^{\sigma-1} - RP] \Biggr\} \ ,
\end{eqnarray}
\begin{eqnarray}
\Delta R(j) &=& x {(\sigma-2) \over \sigma } \Biggl\{
R_{k-4}^{\sigma-3} P P^{\sigma-1} [1+R]
+ R_{k-3}^{\sigma-3} [1-P ]
P^{\sigma-1} [1+R] \nonumber \\ &&
+ R_{k-3}^{\sigma-3} P [1 - P^{\sigma-1} 
- RP^{\sigma-1}]
+ R_{k-2}^{\sigma-3}[1-P][1-P^{\sigma-1}
- RP^{\sigma-1}] \Biggr\} \ .
\end{eqnarray}
Also
\begin{eqnarray}
\Delta R(k) = - R^5 p^4 (\sigma-1)^2 \ .
\end{eqnarray}
So, in all from Fig. \ref{D7} we get
\begin{eqnarray}
\Sigma_{y=a}^k \Delta R(y)
&=& {x(\sigma-2) \over \sigma} PR^3 (R_{k-2}^{\sigma-2} -
R_{k-3}^{\sigma-2})
+ {x (\sigma-2) R_{k-4}^{\sigma-3} \over 2 \sigma} [
-3R^2P^2 + 3R^2P^{\sigma-1} + R^3]- x R^5
\nonumber \\ && \ +
{x (\sigma-2) \over 2 \sigma} R_{k-3}^{\sigma-3} [
6R^2 P^2 - 6R^2 P^{\sigma-1} ]
+ {x(\sigma-2) \over 2 \sigma} R_{k-2}^{\sigma-3} [
3R^2(P^{\sigma-1}-P^2) - R^3] \ . 
\label{INALLA} \end{eqnarray}

\begin{figure}
\centerline{\psfig{figure=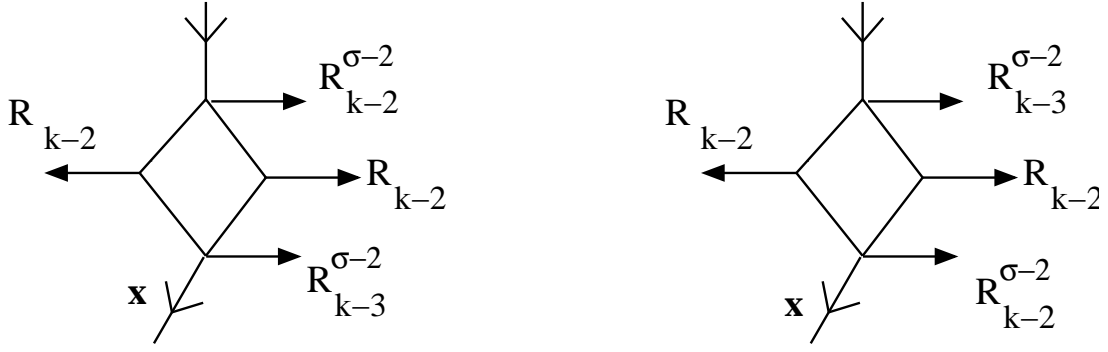}}
\vspace{0.3 in} \caption{Diagrams for Eq. (\protect{\ref{RAEQ}}).
In the left-hand diagram the bottom site will be culled if
the bond "x" is culled.  In that case the entire cluster
will be culled.  So the stability of this cluster depends on
the state of bonds below "x," and therefore this diagram can be
extended.  In contrast, the right-hand cluster survives
culling no matter what may be the state of bond "x."
Hence this diagram truncates and we discard it.}
\label{D10} \end{figure}

We next consider the contributions from the diagrams of Fig.
\ref{D8}.  We get
\begin{eqnarray}
\Delta R (a) &=& {x(\sigma-1)\over 2 \sigma} \Biggl\{
2R (R_{k-2}^{\sigma-2})^2
[1-P^{\sigma-1}-R] +2 R P^{\sigma-1}
(R_{k-3}^{\sigma-2})^2 \nonumber \\ &&
+ R^2 (R_{k-3}^{\sigma-2})^2
+ R^2R_{k-2}^{\sigma-2}R_{k-3}^{\sigma-2} \Biggr\} \ .
\label{RAEQ} \end{eqnarray}
In Fig. \ref{D10} we show the evaluation of the last term in this
result.  Note that for a diagram with a loop, it can matter
which way the diagram is entered. 

\vspace{0.3 in}

\begin{eqnarray}
\Delta R(b) &=& -2{\sigma-1 \over \sigma} x R^2 \Biggl( R_{k-3}^{\sigma-2}
[P^{\sigma-1} + RP] + R_{k-2}^{\sigma-2}
[1-P^{\sigma-1} - R P ] \Biggr) \ ,
\end{eqnarray}
\begin{eqnarray}
\Delta R(c) &=& - x{(\sigma-1) \over \sigma}
R[R_{k-3}^{\sigma-2}P + R_{k-2}^{\sigma-2}(1-P)]^2 \ ,
\end{eqnarray}
\begin{eqnarray}
\Delta R(d) &=& - 2 {x (\sigma-1) \over \sigma}
RR_{k-2}^{\sigma-2} [R_{k-3}^{\sigma-2}P^{\sigma-1}
+ R_{k-2}^{\sigma-2} (1-P^{\sigma-1})] \ .
\end{eqnarray}
\begin{eqnarray}
\Delta R(e) &=& 2x {\sigma-1 \over \sigma} R^2[R_{k-3}^{\sigma-2}P^{\sigma-1}
+ R_{k-2}^{\sigma-2} (1-P^{\sigma-1})] \ .
\end{eqnarray}
\begin{eqnarray}
\Delta R(f) &=& 2{x(\sigma-1) \over \sigma}
RR_{k-2}^{\sigma-2} [R_{k-3}^{\sigma-2}P
+ R_{k-2}^{\sigma-2} (1-P)] \ .
\end{eqnarray}
So, in all we get from the diagrams of Fig. \ref{D8}
\begin{eqnarray}
\Sigma_{y=a}^f \Delta R(y) &=& -2x {\sigma-1 \over \sigma} 
[R_{k-3}^{\sigma-2}-R_{k-2}^{\sigma-2}]
R^3P + {x(\sigma-1)R \over \sigma}[P^{\sigma-1} - P^2]
[R_{k-2}^{\sigma-2} - R_{k-3}^{\sigma-2}]^2 \nonumber \\ &&
+ {x(\sigma -1) \over 2 \sigma} R^2 
[ R_{k-3}^{\sigma -2} - R_{k-2}^{\sigma -2} ]
[ R_{k-3}^{\sigma -2} + 2 R_{k-2}^{\sigma -2} ] \ . 
\label{INALLB} \end{eqnarray}

\begin{figure}
\centerline{\psfig{figure=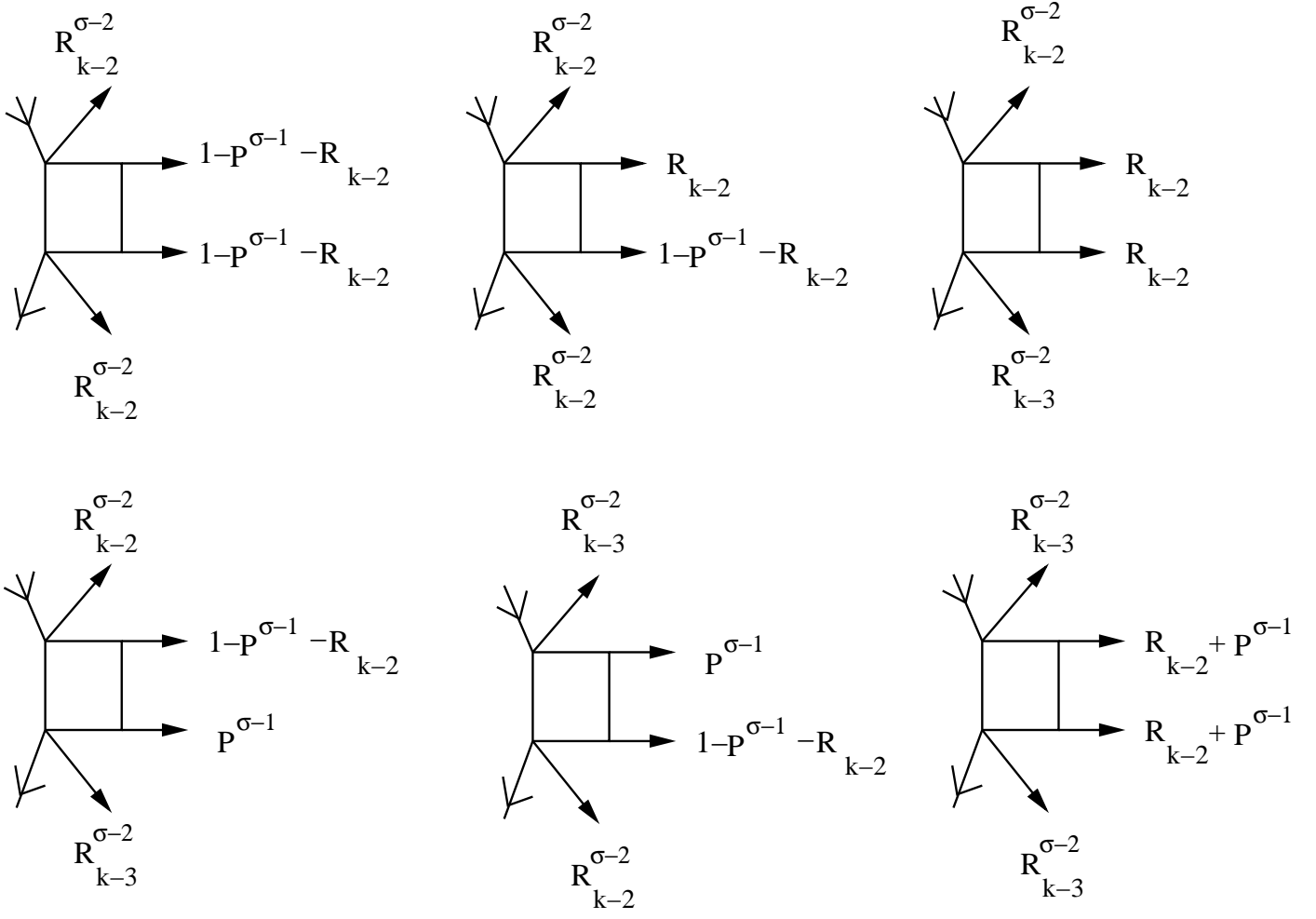}}
\vspace{0.3 in} \caption{Evaluation of $\Delta R(a)$ of Eq.
(\protect{\ref{SQ5EQ}}).}
\label{SQ5FIG} \end{figure}

For the diagrams of Fig. \ref{D9} we have
\begin{eqnarray}
\Delta R(a) &=& { x (\sigma-1) \over \sigma } \Biggl\{  
R^2 R_{k-2}^{\sigma-2} R_{k-3}^{\sigma-2}
+ (R_{k-3}^{\sigma-2})^2 (R+P^{\sigma-1})^2
\nonumber \\ &&
+ 2 R_{k-2}^{\sigma-2} R_{k-3}^{\sigma-2} P^{\sigma-1}
(1-P^{\sigma-1} - R)
\nonumber \\ && + (R_{k-2}^{\sigma-2})^2 (1-P^{\sigma-1}-R)
(1-P^{\sigma-1}+ R) \Biggr\} \ ,
\label{SQ5EQ} \end{eqnarray}
as is illustrated in Fig. \ref{SQ5FIG}.

\vspace{0.3 in}

For the other diagrams in Fig. \ref{D9} we find
\begin{eqnarray}
\Delta R(b) &=& - x {(\sigma-1) \over \sigma}
(R_{k-2}^{\sigma-2})^2 R^2 \ ,
\end{eqnarray}
\begin{eqnarray}
\Delta R(c) &=& + 2x R^3R_{k-2}^{\sigma-2}  \ .
\end{eqnarray}
\begin{eqnarray}
\Delta R(d) &=& - 2x R^3[ R_{k-2}^{\sigma-2} (1-P)
+ R_{k-3}^{\sigma-2}  P] \ ,
\end{eqnarray}
\begin{eqnarray}
\Delta R(e)  &=&  -2x R \Biggl( R_{k-2}^{\sigma-2}
[1-P^{\sigma-1}(1 + R) - R^2 P] \nonumber \\ &&
+ R_{k-3}^{\sigma-2} [ P^{\sigma-1}(1 + R) + P R^2 ] \Biggr) \ .
\end{eqnarray}
\begin{eqnarray}
\Delta R(f) &=& -2x {\sigma -1 \over \sigma}
\Biggl[ R_{k-2}^{\sigma-2} (1-P) + R_{k-3}^{\sigma-2} P \Biggr]
\Biggl[ R_{k-2}^{\sigma-2} [1-P^{\sigma-1} - R P] 
\nonumber \\ && + R_{k-3}^{\sigma-2} [P^{\sigma-1} +
R P] \Biggr] \ ,
\end{eqnarray}
\begin{eqnarray}
\Delta R(g) &=& -2x {(\sigma-1) \over \sigma} R_{k-2}^{\sigma-2} \Biggl[
R_{k-2}^{\sigma-2} [1-P^{\sigma-1} - R P^{\sigma-1}]
+ R_{k-3}^{\sigma-2} P^{\sigma-1}[1 + R] \Biggr] \ ,
\end{eqnarray}
\begin{eqnarray}
\Delta R(h) &=& 2x {\sigma -1 \over \sigma} R_{k-2}^{\sigma-2} \Biggl[
R_{k-2}^{\sigma-2} [1-P^{\sigma-1} - R P]
+ R_{k-3}^{\sigma-2} [P^{\sigma-1} + RP] \Biggr] \ ,
\end{eqnarray}
\begin{eqnarray}
\Delta R(i) &=& 2x R \Biggl[ R_{k-2}^{\sigma-2} [1-P^{\sigma-1}
- R P^{\sigma-1}] + R_{k-3}^{\sigma-2} [1+ R]P^{\sigma-1} \Biggr] \ ,
\end{eqnarray}
\begin{eqnarray}
\Delta R(j) &=& - x {(\sigma-1) \over \sigma}
[R_{k-2}^{\sigma-2}(1-P^{\sigma-1}) + R_{k-3}^{\sigma -2} P^{\sigma-1}]^2 \ .
\end{eqnarray}
\begin{eqnarray}
\Delta R(k) &=& 2x {\sigma -1 \over \sigma}
[R_{k-2}^{\sigma-2}(1-P^{\sigma-1}) + R_{k-3}^{\sigma -2} P^{\sigma-1}]
[R_{k-2}^{\sigma-2}(1-P) + R_{k-3}^{\sigma -2} P] \ .
\end{eqnarray}
So, in all from the diagrams of Fig. \ref{D9} we get
\begin{eqnarray}
\Sigma_{y=a}^k \Delta R(y)
&=& 4xR^3 P[R_{k-2}^{\sigma -2} - R_{k-3}^{\sigma -2} ]
+ x {\sigma -1 \over \sigma} \Biggl[ (R_{k-2}^{\sigma-2})^2
[-2R^2 +2R(P^{\sigma -1} - P^2)] \nonumber \\ && \
+ (R_{k-3}^{\sigma -2})^2 [ R^2 + 2R(P^{\sigma -1} -P^2)]
+ R_{k-2}^{\sigma -2} R_{k-3}^{\sigma -2} [R^2 -4R(P^{\sigma -1} - P^2)
] \Biggr] \ .
\label{INALLC} \end{eqnarray}

\subsection{$k$-Core Percolation, Hexagons}

Now consider the diagrams of Fig. \ref{H1a}, but for the
susceptibility instead.  Here and below the vertices (or vertex) which are part
of the chain are indicated by filled circles. With $y=2\sigma^3p^6$
we have
(with $R \equiv R_{k-2}$)
\begin{eqnarray}
\Delta R(a) &=& y \Biggl\{ R_{k-4} [P^2 (1+2R+3R^2 +4R^3)
+5R^4P +R^5] \nonumber \\ &&
+ 2R_{k-3}P \Biggl[ 1-R^4 -P(1+R+R^2+R^3)
+R(1-R^3) -RP(1+R+R^2) \nonumber \\ &&
+R^2(1-R^2) -R^2P(1+R) 
+R^3(1-R) -R^3P \Biggr] \nonumber \\ &&
+R \Biggl[ (1-P-R)^2 (1+2R+3R^2+4R^3) +5R^4(1-P -R)
\Biggr] \Biggr\} 
\end{eqnarray}
\begin{eqnarray}
\Delta R(b) &=& -2y \Biggl\{ R_{k-3}P (1+R+R^2+R^3+R^4+R^5) 
\nonumber \\ && +R [1 -P (1+R+R^2+R^3+R^4+R^5)] \Biggr\} 
\end{eqnarray}
\begin{eqnarray}
\Delta R (c) &=& -2y \Biggl\{ R_{k-4}P^2 (1+R+R^2+R^3+R^4)
+R_{k-3}P (1-P) (1+R+R^2+R^3+R^4) \nonumber \\ &&
+R_{k-3}P [1 -P (1+R+R^2+R^3+R^4)] +
\nonumber \\ && R(1-P)[1-P (1+R+R^2+R^3+R^4)]
\Biggr\} \end{eqnarray}
\begin{eqnarray}
\Delta R(d) &=& -2y \Biggl\{  R_{k-4} P^2(1+R)(1+R+R^2+R^3)
+R_{k-3} P (1+R)[1-P(1+R+R^2+R^3)] \nonumber \\ &&
+ R_{k-3} P [1-P(1+R)](1+R+R^2+R^3) \nonumber \\ &&
+ R [1-P(1+R)][1-P(1+R+R^2+R^3)] \Biggr\} 
\end{eqnarray}
\begin{eqnarray}
\Delta R(e) &=& -y \Biggl\{  R_{k-4} P^2(1+R+R^2)^2
+2R_{k-3} P (1+R+R^2)[1-P(1+R+R^2)] \nonumber \\ &&
+ R [1-P(1+R+R^2)]^2 \Biggr\} 
\end{eqnarray}
\begin{eqnarray}
\Delta R(f) &=& 2y \Biggl\{ R_{k-3} P (1+R+R^2+R^3+R^4)]
\nonumber \\ && + R [1-P(1+R+R^2+R^3+R^4)] \Biggr\} 
\end{eqnarray}
\begin{eqnarray}
\Delta R(g) &=& 2y \Biggl\{ R_{k-4}P^2 (1+R+R^2+R^3)
+R_{k-3}P (1-P) (1+R+R^2+R^3) \nonumber \\ &&
+R_{k-3}P [1 -P (1+R+R^2+R^3)] 
\nonumber \\ && + R(1-P)[1-P(1+R+R^2+R^3)] \Biggr\}
\end{eqnarray}
\begin{eqnarray}
\Delta R(h) &=& 2y \Biggl\{ R_{k-4}P^2 (1+R)(1+R+R^2)
+R_{k-3}P [1-P(1+R)] (1+R+R^2) \nonumber \\ &&
+R_{k-3}P(1+R) [1 -P (1+R+R^2)] \nonumber \\ &&
+ R[1-P(1+R)][1-P(1+R+R^2)] \Biggr\} \ . 
\end{eqnarray}
In addition we have the analog of diagram (k) of Fig. \ref{D7}:
\begin{eqnarray}
\Delta R(i) &=& -2y R^7 \ ,
\end{eqnarray}
We call the sum of these contributions $\Gamma_{H,A}$.  We have
\begin{eqnarray}
\Gamma_{H,A} &=&
2 \sigma^3 p^6 \Biggl[ R_{k-4}[-5P^2 R^4 + 5P R^4 + R^5]
+2 R_{k-3}P[ 5R^4P -5R^4 -R^5]-5R^5P^2
\nonumber \\ && +(5R^5+2R^6) P -R^6 -2R^7 \Biggr] \ .
\end{eqnarray}

\vspace{0.3 in}
\begin{figure}
\centerline{\psfig{figure=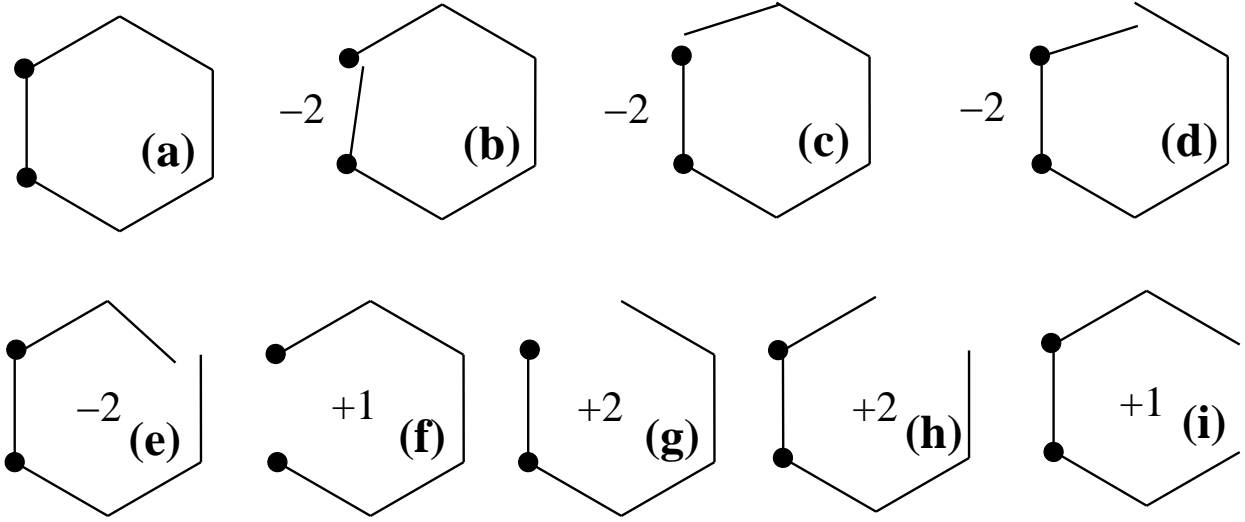}}
\vspace{0.3 in} \caption{Susceptibility hexagons on either H or B lattices.}
\label{H2} \end{figure}

Now we consider the diagrams of Fig. \ref{H2}.  We have
\begin{eqnarray}
\Delta R(a) &=& 2y \Biggl\{ R_{k-3}^2 [P^2 (1+2R+3R^2)
+ 4P R^3 +R^4] \nonumber \\ &&
+ 2 R_{k-3}R [P(1-P-R)(1+2R+3R^2)]
+ R_{k-3} R^5 \nonumber \\ &&
+R^2 [(1-P-R)^2 (1+2R+3R^2) + 4(1-P-R) R^3] \Biggr\} 
\end{eqnarray}
\begin{eqnarray}
\Delta R(b) &=& -4yR^5[ R(1-P)+ R_{k-3}P]
\end{eqnarray}
\begin{eqnarray}
\Delta R(c) &=& -4y \Biggl\{ RR_{k-3} [(1+R+R^2+R^3+R^4)P]
+R^2[1-P(1+R+R^2+R^3+R^4)] \Biggr\}
\end{eqnarray}
\begin{eqnarray}
\Delta R(d) &=& -4y \Biggl[ R(1-P)+R_{k-3}P \Biggr]
\nonumber \\ && \ \times \Biggl[
R[1-P(1+R+R^2+R^3)] + R_{k-3}P(1+R+R^2+R^3)\Biggr]
\end{eqnarray}
\begin{eqnarray}
\Delta R(e) &=& -4y \Biggl[ R[1-P(1+R)] + R_{k-3}P(1+R)\Biggr]
\nonumber \\ && \ \times
\Biggl[ R[1-P(1+R+R^2)] + R_{k-3} P(1+R+R^2)\Biggr]
\end{eqnarray}
\begin{eqnarray}
\Delta R(f) &=& 2yR^6
\end{eqnarray}
\begin{eqnarray}
\Delta R(g) &=& 4y \Biggl\{ RR_{k-3} [(1+R+R^2+R^3)P]
+R^2[1-P(1+R+R^2+R^3)] \Biggr\}
\end{eqnarray}
\begin{eqnarray}
\Delta R(h) &=& 4y \Biggl[ R(1-P)+R_{k-3}P \Biggr] \Biggl[
R[1-P(1+R+R^2)] + R_{k-3}P(1+R+R^2)\Biggr]
\end{eqnarray}
\begin{eqnarray}
\Gamma_i &=& 2y \Biggl[ R[1-P(1+R)] + R_{k-3}P(1+R)\Biggr]^2
\end{eqnarray}
We call the sum of all these contributions $\Gamma_{H,B}$. We have
\begin{eqnarray}
\Gamma_{H,B} &=& 4 \sigma^3 p^6 
\Biggl\{ R_{k-3}^2 \Biggl( 4R^3[P-P^2] +R^4 \Biggr)  
+R_{k-3} \Biggl[ 8R^4 (P^2 - P ) -4R^5P
+ R^5 \Biggr] \nonumber \\ &&
-4R^5P^2 +4R^5P +4P R^6  -2 R^6 \Biggr\}  \ .
\end{eqnarray}

\vspace{0.3 in}
\begin{figure}
\centerline{\psfig{figure=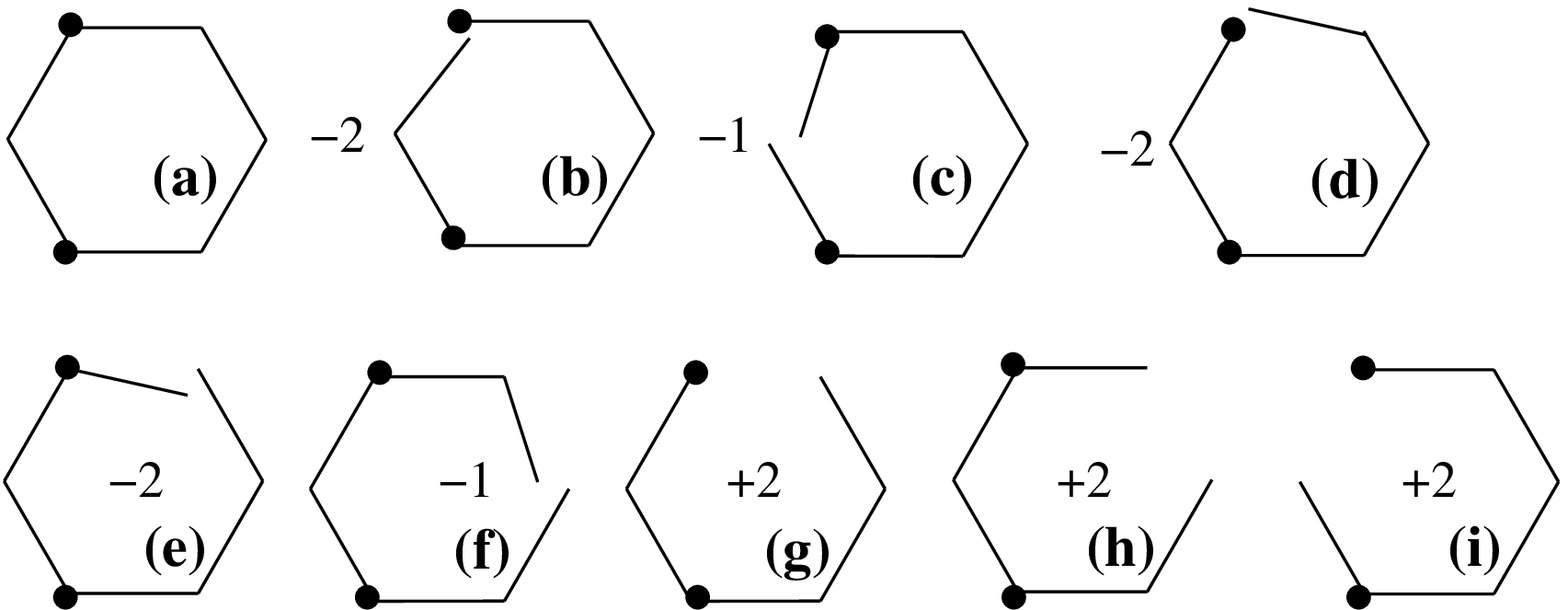}}
\vspace{0.3 in} \caption{Susceptibility hexagons on either H or B lattices.}
\label{H3} \end{figure}

Next we consider the diagrams of Fig. \ref{H3}.
\begin{eqnarray}
\Delta R(a) &=& 2y \Biggl\{ R_{k-3}^2 [ R^4 + 4R^3P + 2 R^2 P^2 
+ R P^2 ] \nonumber \\ &&
+RR_{k-3}[2RP(1-P-R) +4R^2P (1-P-R)+R^4]
\nonumber \\ && + R^2[R(1-P-R)^2 + 2R^2(1-P-R)^2
+ 4R^3(1-P-R)] \Biggr\} 
\end{eqnarray}
\begin{eqnarray}
\Delta R(b) &=&-4y R^4 \Biggl[ R[1-P(1+R)]
+ R_{k-3} P(1+R) \Biggr]
\end{eqnarray}
\begin{eqnarray}
\Delta R(c) &=& - 2y R^3 [ R(1-P)+R_{k-3}P]^2
\end{eqnarray}
\begin{eqnarray}
\Delta R(d) &=& - 4y R^2 \Biggl[ R[1-P(1+R+R^2+R^3)]+R_{k-3}P
(1+R+R^2+R^3)]\Biggr]
\end{eqnarray}
\begin{eqnarray}
\Delta R(e) &=& - 4yR \Biggl[ R(1-P) +R_{k-3}P \Biggr]
\Biggl[ R[1-P(1+R+R^2)] + R_{k-3}P(1+R+R^2)] \Biggr]
\end{eqnarray}
\begin{eqnarray}
\Delta R(f) &=& - 2yR \Biggl[ R[1-P(1+R)] + R_{k-3}P(1+R)\Biggr]^2
\end{eqnarray}
\begin{eqnarray}
\Delta R(g) &=& 4yR^2 \Biggl[ R[1-P(1+R+R^2)]
+ R_{k-3} P(1+R+R^2) \Biggr]
\end{eqnarray}
\begin{eqnarray}
\Delta R(h) &=& 4yR \Biggl[ R[1-P(1+R)] + R_{k-3} P(1+R) \Biggr]
\Biggl[ R(1-P) + R_{k-3} P \Biggr]
\end{eqnarray}
\begin{eqnarray}
\Delta R(i) &=& 4yR^4 [ R(1-P) + R_{k-3} P]
\end{eqnarray}
We call the sum of these contributions $\Gamma_{H,C}$.  We find that
$\Gamma_{H,C}=\Gamma_{H,B}$.

\vspace{0.3 in}
\begin{figure}
\centerline{\psfig{figure=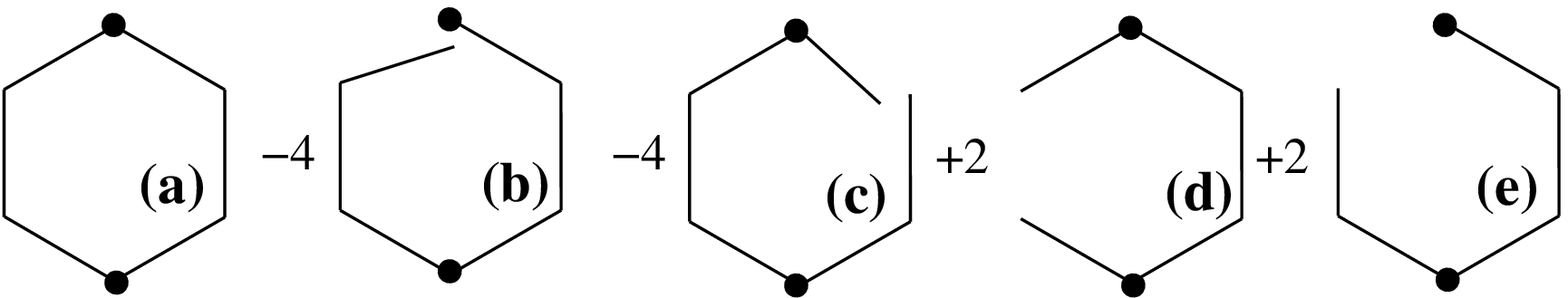}}
\vspace{0.3 in} \caption{Susceptibility hexagons on either H or B lattices.}
\label{H4} \end{figure}

Next we consider the diagrams of Fig. \ref{H4}.
\begin{eqnarray}
\Delta R(a) &=& y\Biggl\{ R_{k-3}^2[R^4+4R^3P
+ 2R^2 P^2] + RR_{k-3}[ 4R^2P (1-P-R) + R^4]
\nonumber \\ && +R^2[2(1-P-R)^2R^2 + 4 (1-P-R)R^3] \Biggr\}
\end{eqnarray}
\begin{eqnarray}
\Delta R(b) &=& -4yR^3 \Biggl[ R[1-P(1+R+R^2)]
+ R_{k-3} P(1+R+R^2)
\Biggr] 
\end{eqnarray}
\begin{eqnarray}
\Delta R(c) &=& -4yR^2 \Biggl[ R[1-P(1+R)] + R_{k-3} P(1+R)
\Biggr] [R(1-P) + R_{k-3}P]
\end{eqnarray}
\begin{eqnarray}
\Delta R(d) &=& 2yR^2 [R(1-P) + R_{k-3}P]^2
\end{eqnarray}
\begin{eqnarray}
\Delta R(e) &=& 4yR^3 \Biggl[ R[1-P(1+R)] + R_{k-3} P(1+R)
\Biggr] \ .
\end{eqnarray}
We call the sum of these contributions $\Gamma_{H,D}$.  We have
$\Gamma_{H,D} = \Gamma_{H,B}/2$. Then the total contribution
to the renormalization of $R$ from hexagons is
\begin{eqnarray}
\Delta R_{\rm Hex}  &=& 2p^6 \sigma^3 \Biggl\{
R_{k-4} \Biggl[ 5R^4(P-P^2)+R^5\Biggr] + R_{k-3}^2 \Biggl[
20R^3(P-P^2) +5R^4 \Biggr] -25 R^5(P^2 -P) \nonumber \\ && 
+ R_{k-3} \Biggl[ 50 (P^2-P)R^4 - 22 PR^5 + 5 R^5 \Biggr]
+ 22 R^6P - 11R^6 -2R^7 \Biggr\} \ .
\label{HEXEQ} \end{eqnarray}
\section{Conclusion}

We now have all that we need to evaluate whether or not the perturbed 
ratio $r$, i.e. the right-hand side of Eq. (\ref{ratioeq}), is unity or not to order $1/\sigma^3$.
Using Eqs. (\ref{HEXEQ}) and (\ref{PHIHEX}) we see that the
contributions from the hexagons do not contribute to the rhs of
Eq. (\ref{ratioeq}).  Furthermore, using Mathematica~\cite{WOLF}, we summed the contributions from the squares
in Eqs. (\ref{PHISQEQ}), (\ref{INALLA}), (\ref{INALLB}), and
(\ref{INALLC}), which led to the result
(at the transition in the EOS) that
\begin{eqnarray}
r = 1 + (3 \sigma p^4/2) R^2[R-R_{k-3}][(2-k)R + \sigma Q R_{k-3}]
+ {\cal O} (\sigma^{-4} ) \ .
\label{RESULT} \end{eqnarray}
In writing this result we have assumed that the effect of
$1/d$ perturbations is merely to shift the location of the
pole in the susceptibility.  It is not obvious that this is
a correct assumption, since we do not know {\it a priori}
that for finite $d$ the suceptibility diverges (as it does
in the other problems for which we cited the use of the $1/d$
expansion)~\cite{MEF}. A discussion is given in Appendix B to
justify this assumption.

Since $p \sim 1/\sigma$, the term proportional to $p^4$ is of order
$\sigma^{-3}$.  Note that to leading order in $1/\sigma$ we have that 
\begin{eqnarray}
R_m &=& (\sigma Q)^m (m!)^{-1} e^{- \sigma Q} \ ,
\label{REQ} \end{eqnarray}
which implies that $(2-k)R + \sigma Q R_{k-3} = {\cal O} (1/\sigma)$,
so that to order $(1/\sigma)^3$, the result of Eq. (\ref{RESULT})
is that $r=1$, which means that to this order the singularity in the EOS
and the divergence in the susceptibility coincide.  Note that
this result is not a trivial  one in that it implicitly involves
Eq. (\ref{REQ}).  Also, it can hardly be a coincidence that the
subtleties of diagram counting lead to this result.  It would not
totally surprise us if this result could be obtained to all
orders in $1/\sigma$ by some type of
Ward identity, even in the absence of a field theoretical formulation.

The coincidence of the two singularities suggests that the unusual nature 
of the $k\ge3$-core transition survives in large, but finite, spatial 
dimensions.  Given the absence of numerical confirmation of such 
a transition in simple isotropic $k$-core 
models on square, cubic, and triangular
lattices, our results motivate further numerical studies of models that
are more mean field-like in the sense that the range (number) of
nearest neighbors $\sigma+1$ is larger than previous models.  Of course,
to compare with our results one would need to study hypercubic lattices
at large $d$.  Since ``culling'' leads to nonlocal truncations, it will not
be easy to  simulate systems at large $d$ to check our work.
However, should such a program be undertaken, one might want
to have actual expressions for the shift in the critical point
and the shift in the jump of the order parameter due to finite
dimensionality.  To get the shift in $p_c$, one needs to
evaluate the rhs of Eq. (\ref{DELTAP}), using Eqs. (\ref{PHISQEQ})
and (\ref{PHIHEX}).  This result will include all corrections
up to and including order $1/ \sigma^{3}$.  However, it is necessary
to expand the quantities such as $p_0$, $Q_0$, and $R_m$
in powers of $1/\sigma$, as is indicated by Eqs. (\ref{PZERO})
and (\ref{RMEQ}).  Furthermore, to get the shift in the jump in the order
parameter involves evaluating the rhs of Eq. (\ref{DELTAQ}) and
for this we had recourse to Mathematica~\cite{WOLF}.

\begin{figure}
\centerline{\psfig{figure=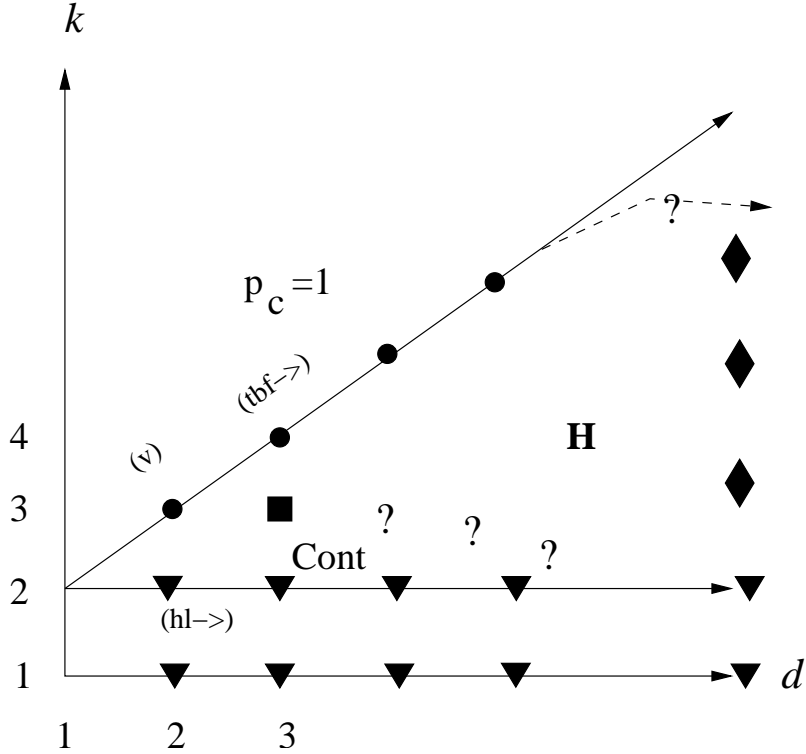}}
\vspace{0.3 in} \caption{Summary of known results for $k$-core percolation on the $d$-dimensional
hypercubic lattice.  Triangles represent ordinary percolation
for $k=1$ and biconnectedness for $k=2$ (whose critical exponents
are governed by the fixed point of ordinary percolation~\protect\cite{BICON}).
Circles represent systems for which $p_c$ was shown to be unity
for $d=2$ by van Enter~\cite{ACDvanE} and by Toninelli {\it et al}.~\cite{TBF} for larger $d$.
The filled square represents the result of Ref.~\onlinecite{KL} for $d=k=3$ 
where the transition appears to be continuous.  
This result therefore suggests the existence
of a region labeled "Cont" of a continuous transition with a critical
point probably similar to that of ordinary percolation. The
filled diamonds represent the present work suggesting that the hybrid
transition survives for large $d$ (but with $k$ not comparable to $d$).  The
question marks indicate that the
boundaries between the continuous transition (filled square) and
the hybrid transition, as well as that between $p_c=1$ and the hybrid
regime, are unclear at present.} 
\label{CUBIC} \end{figure}

In order to assess the implications of our work when placed in context
with known results we refer to the phase diagram in the $k$-$d$ plane
in Fig. \ref{CUBIC}
where we indicate existing results for $k$-core percolation.
The two regimes which are most securely established are a)
for $k=2$ the model has~\cite{BICON} the same critical point as ordinary
percolation and b) for $k \geq d+1$ the critical concentration for
percolation has been shown~\cite{TBF} to be $p_c=1$.  The present work suggests that the hybrid transition found for
$d=\infty$ persists into the regime $k \ll d < \infty$.  In addition, numerical work indicates that the transition may be 
continuous for $k=d=3$~\cite{KL}. Since this is the only data point which gives a continuous transition
for $k>2$, it is important to confirm this result on larger samples,
although our discussion will assume the validity of this result.
The $1/d$ result of the present work indicates that the hybrid
transition occurs for $k \geq 3$ at large $d$, so it is clear that the
continuous transition seen at $k=d=3$ must disappear as $d$ is increased 
for $k=3$.  The details of this boundary between
hybrid and continuous transitions is unclear at present, except that
it seems almost certain that $k=3$ is essentially different from $k=2$. 
Furthermore, at present there is no evidence of yet another phase between the
hybrid phase and the $p_c=1$ phase at high $d$.  This phase
boundary might remain at $k=d+1$.  Alternatively, since finite-dimensional
fluctuations reduce $p_c$ in the hybrid phase from its mean-field
or Bethe lattice value, it is possible that the phase boundary between
these two phase falls below the line $k=d+1$, as indicated
by the question mark in Fig. \ref{CUBIC}.  From the perspective of Fig.
\ref{CUBIC} it would be interesting to develop a realization of
$k$-core percolation for noninteger $k$.

It is interesting to note how the $1/\sigma$ expansion for $k$-core
percolation compares to that for other systems.  For self-avoiding
walks~\cite{ISING}, the Ising model~\cite{ISING}, spin glasses~\cite{SINGH},
and for percolation~\cite{HDPERC}, the expansions for the critical value
of the coupling constant involves coefficients of $1/\sigma^n$, which are
rational fractions (i. e. the ratio of two finite integers).  In
contrast, for lattice animals one sees the appearance of the transcendental
number $e$ in the coefficients of $1/\sigma^n$~\cite{ABH}.  From Eq. (20)
one sees a similar result for $k$-core percolation.  It has been
noted~\cite{SLC} that the unusual mean-field value of the
correlation length exponent for $k$-core percolation is identical to
that for lattice animals~\cite{ANIMALS}. Thus, it is not surprising that
these two models show similar unusual characteristics in their
$1/\sigma$ expansion for the critical coupling constant.

\section{Discussion}

The model of $k$-core percolation sets a constraint on the number of 
occupied neighboring bonds (or sites in the corresponding site problem). 
So the physics of $k$-core pertains to systems with 
nontrivial constraints such 
as the onset of long-range orientational order in solid ortho-para ${\rm H}_2$ 
mixtures and the onset of rigidity in a mechanical system.  
Given the equivalence 
between rigidity percolation and $(g+1)$-core percolation on the Bethe lattice \cite{RIGIDCORE}, 
the results here suggest that provided $g$ is at least $2$ in high dimensions, 
the percolation transition should be of a hybrid nature where there is 
a jump in the order parameter accompanied with a diverging length scale.  
We note that there exists a field theory~\cite{OBUKHOV} of rigidity
percolation that exhibits a purely first-order transition,
which is very different from the hybrid nature of $k$-core
percolation. Given our results, as well as the results of SLC, it seems 
that this task of constructing a field theory for rigidity percolation should 
be revisited. 

Other systems with nontrivial constraints include the dynamics 
of an interface separating two magnetic domains in a random field Ising 
model~\cite{JR}.  When the disorder is strong, it is likely that more than 
several neighbors 
must have flipped previously in order to propagate the interface, again,  
giving rise 
to $k$-core physics. However, finite-dimensional 
simulations of such interfaces have only uncovered ordinary percolation exponents so far ~\cite{JR}, as opposed to some sort of hybrid transition.

While not directly related to $k$-core percolation, spin glasses are 
systems where a set of nontrivial constraints 
cannot all be satisfied.  This property is otherwise known as frustration.
Curiously enough, it turns out that  
the $p\ge3$-spin glass model also exhibits the unusual hybrid 
transition in mean field with 
the same exponents as $k$-core~\cite{KW2}.  It would also be interesting 
to see if a $1/d$ expansion for this system yields the same results 
that we find for $k$-core.  

The physics of glass-forming liquids, a.k.a. the glass transition, 
has received a lot of intense investigation over the years yielding 
a wide range of approaches~\cite{GLASS}.  One approach to modelling 
glassy dynamics is based on 
kinetically constrained spin lattice models, like the Frederickson-Andersen (FA) model~\cite{FA}. In this model, down  
spins on a lattice denote coarse-grained regions of high mobility of the 
liquid, 
while up spins denote regions of low mobility.   A magnetic field favors 
up spins creating large regions of low mobility. As the temperature of 
the system is lowered,  more and more regions of low mobility are 
created eventually leading 
to kinetic arrest of the liquid over large time scales.  An important kinetic constraint on the motion of the spins is   
introduced where a randomly selected spin can flip only if the number 
of neighboring downs spins is equal to or greater than 
some integer whose maximum is 
$\sigma+1$.   This constraint models the caging effect observed 
in glass-forming liquids where particles become trapped by transient 
cages made up of their neighbors. 

 There is an 
{\em exact} mapping between $k$-core percolation and the clustering 
of low mobility regions in the FA model~\cite{RITORT,SBT}.
The mapping to $k$-core is not unexpected given that 
up/down spins can be mapped to occupied/unoccupied sites, and therefore, the  
kinetic constraint maps to the $k$-core condition.  
See Refs.~\onlinecite{RITORT,SBT} for details.  Given this mapping, 
our conclusions apply 
to that type of lattice model, and in high $d$ one should observe a hybrid 
transition at finite temperature, provided $k$ is not comparable to $d$. 
We note that this regime may not be directly relevant for the glass transition 
due to the existence of finite clusters, 
i.e. the cube in $d=k=3$~\cite{clusters}.  Then the density of particles 
in the liquid 
phase is not homogeneous.  However, recent experiments on metallic 
glasses show short-range icosahedral order even in the liquid phase so it 
is not clear if the density should be assumed to be homogeneous, at least 
over some time scale~\cite{KELTON}. Therefore,  
a thorough understanding of $k$-core beyond mean field is of the utmost 
importance to understanding such transitions as the glass transition and 
the onset of orientational ordering in solid ortho-para ${\rm H}_2$ mixtures, and our work seems to 
be the first analytic result to deal with 
its finite-dimensional fluctuations.

\vspace{0.2 in}
ACKNOWLEDGEMENTS  We thank Professor A. J. Liu for very helpful discussions.  JMS was supported
 by NSF-DMR-0087349 and DE-FG02-03ER46087.  

\begin{appendix}
\section{Counting Squares}

Here we consider in detail the counting of squares with
excluded volume corrections which represent corrections
of order $1/\sigma$ relative to leading corrections due
to insertion of squares.
\begin{figure}
\centerline{\psfig{figure=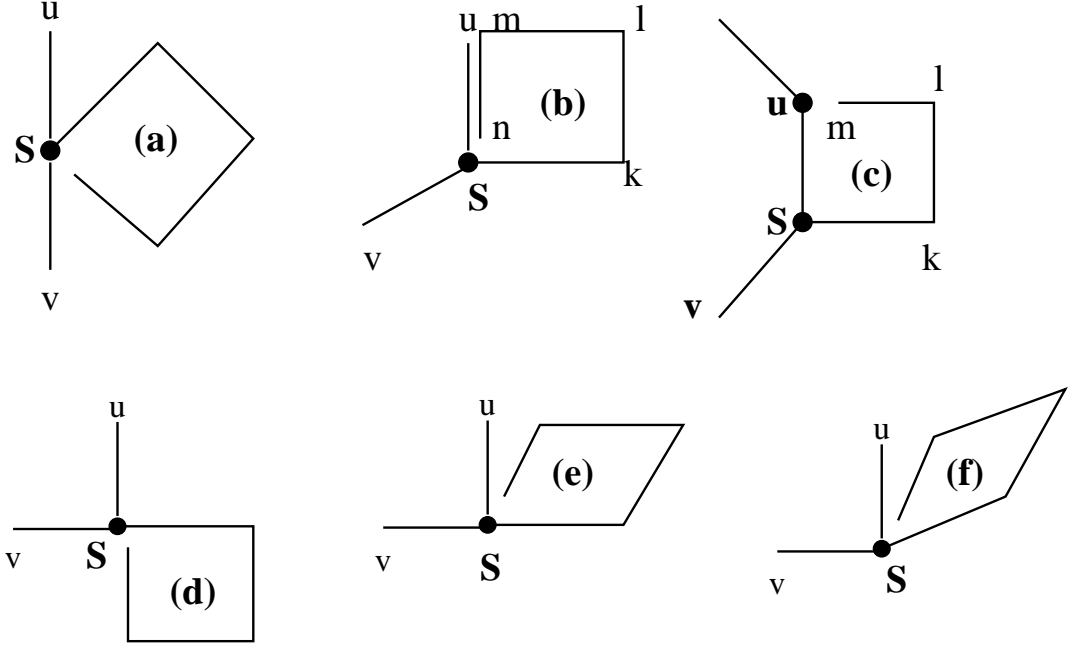}}
\vspace{0.3 in} \caption{Enumeration of the susceptibility diagrams
for the Bethe lattice.
Bonds which are parallel involve the same local displacement and sites
which are close to one another coincide when the diagram is implemented
on the hypercubic lattice.}
\label{SQ8} \end{figure} 
More specifically, we consider the enumeration of pseudo square insertions
at a single vertex, like those shown in Fig. \ref{SQ8}.
Here there are two distinct types of square insertions, all
of which have to subtracted off because their topology is different
on a hypercubic lattice than on a Bethe lattice.  In panel (a)
the situation is simple, in that no sides of the square are
equivalent to the other bonds even on a hypercubic lattice.  The
diagram in panel (b) has to be subtracted off, but note that because
two bonds coincide on the hypercubic lattice, this diagram
is in the class of diagrams of the topology of Fig. \ref{D9},
which here is shown in panel (c).
To avoid subtracting this diagram twice, we do not allow
it in the enumeration of diagrams we wish to subtract off here.
Thus we wish to enumerate pseudo squares (diagrams which would
involve a square on the hypercubic lattice), but only those in which all bonds
are distinct from the bonds of the chain into which the square is inserted.
If the two bonds ($u-S$ and $S-v$) entering the vertex are in the
same direction there are of order $\sigma^2$ ways to form the square.
So the number of configurations of this type is of order $\sigma^2$.
We next count the number of configurations when the
bond $S-v$ is perpendicular to the incoming bond.  There are then
$\sigma-1$ ways to choose this bond.  As shown in panels (d),
(e), and (f), the square can then either (d) be in the same plane as bonds
$u-S$ and $S-v$ (and there are two ways to do this), or (e)
the square can have two bonds parallel to either bond $u-S$ or
bond $S-v$ (there are four ways to do this) and then there are
$\sigma-3$ ways to choose the other bonds of the square to be perpendicular
to bonds $u-S$ and $S-v$, or (f) all bonds of the square can be
perpendicular to the bonds $u-S$ and $S-v$ (there are
$(\sigma -3)(\sigma -5)$ 
way to do this.  So in all, the number of
configurations, $N_s$, of the square and of bond $S-v$ is
\begin{eqnarray}
N_s &=& [\sigma^2 + {\cal O}(\sigma)] +
(\sigma-1) [ 2 + 4 (\sigma -3) + (\sigma -3)(\sigma -5)] \nonumber \\
&=& \sigma^3 [1 + (4/\sigma)] + {\cal O} (\sigma) \ .
\end{eqnarray}
This justifies the prefactors for diagrams (b) and (f) of Fig. \ref{D7}.

\begin{figure}
\centerline{\psfig{figure=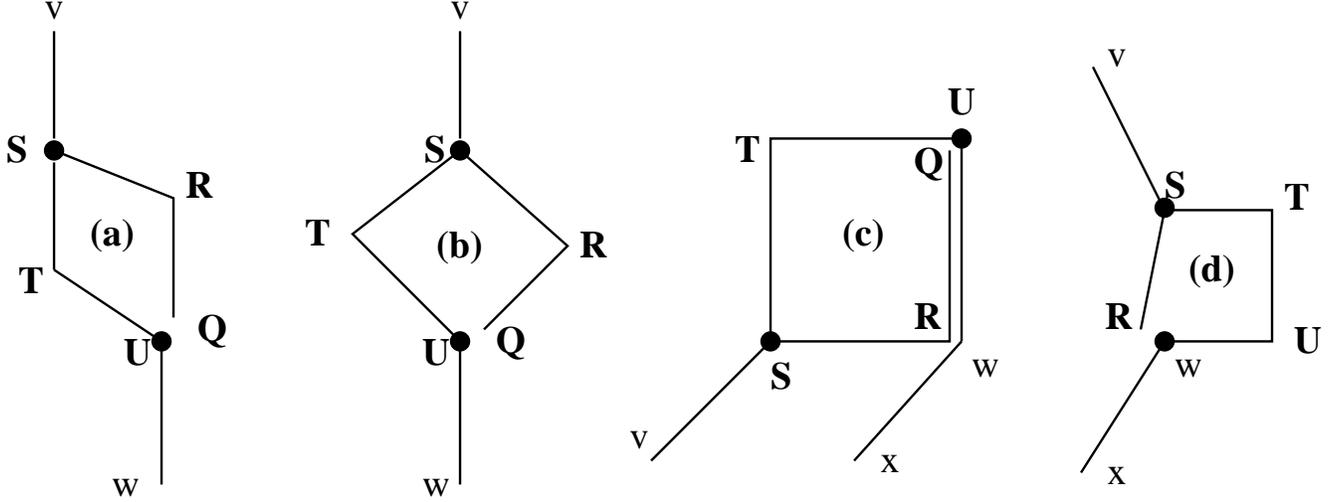}}
\vspace{0.3 in} \caption{Enumeration of the susceptibility diagrams
for the Bethe lattice.
Bonds which are parallel involve the same local displacement and sites
which are close to one another coincide when the diagram is implemented
on the hypercubic lattice.}
\label{SQ9} \end{figure}

The analysis for the diagrams of Fig. \ref{D8} is similar. In panels (a)
and (b) of Fig. \ref{SQ9} we show diagrams which have a
different topology on the Bethe lattice
than on the hypercubic lattice and therefore which must be subtracted off.
In panel (c) we show a special case when bond Q-R coincides with
a bond (U-w) in the chain.  As before, although we do need to subtract off
this diagram, we do not include it in the present enumeration because
it is of the topology of panel (d) which is included in our analysis of
Fig. \ref{D9}.  If bond $S-T$ of panel a is parallel to the
incoming bond then there are $\sigma-1$ ways to complete the square
and $\sigma-1$ ways to add the bond exiting the square.  When bond
$S-T$ of panel b is perpendicular to the incoming bond (there are
$\sigma-1$ ways to do this), there are $\sigma-2$ ways to complete
the square and then $\sigma-1$ ways to add the bond exiting the
square.  Thus in all, $N_s$, the number of ways of configuring the square 
and the exit bond, $U-w$, is
\begin{eqnarray}
N_s &=& (\sigma-1)^2 + (\sigma-1)(\sigma-2)(\sigma-1) \nonumber \\
&=& \sigma^3[1-(3/\sigma)] + {\cal O}(\sigma) \ .
\end{eqnarray}
This calculation justifies the prefactors for diagrams (b) and (e) of
Fig. \ref{D8}.

The diagrams of Fig. \ref{D9} are more straightforward because for
them one does not have to consider the possibility of double
counting the subtractions.

\section{Shift of Singularity in the Susceptibility}
 
In Sec. VI we evaluated the ratio $r$ in the series for the susceptibility
and interpreted the result as giving a shift in the singularity.  Here
we justify this interpretation.  Note that the vertex renormalization
(in which the unperturbed ratio $r_0$ is replaced by
$r_0+ r_0 \Delta $) gives the susceptibility as
\begin{eqnarray}
\chi = \sum_n r_0^n [1 + n \Delta ] \ .
\end{eqnarray}
In writing this result we noted that for a chain of $n$ bonds
the vertex renormalization could
be placed at any one of order $n$ vertices.  Of course, there are
end effects, so that really, if we include end effects, we would write
\begin{eqnarray}
\chi = \sum_n r_0^n [1 + n \Delta + \epsilon ] \ ,
\label{END1} \end{eqnarray}
where both $\Delta$ and $\epsilon$ include all contributions of order
$\sigma^{-2}$ and $\sigma^{-3}$.  Now we consider the contributions
of {\it two} vertex renormalizations.  If the two vertices do not
interfere with one another then their contribution is $r_0\Delta^2$.  For
a chain of $n$ bonds, there are of order $n^2/2$ such configurations.
So in analogy with Eq. (\ref{END1}), these contributions are of the form
\begin{eqnarray}
\delta \chi &=& \sum_n r_0^n [ n^2 \Delta^2 /2 + n \epsilon + \eta ]\ ,
\end{eqnarray}
where $\epsilon$ comes from configurations involving two squares or
higher order configurations, none of which were counted up to
order $\sigma^{-3}$ in Eq. (\ref{END1}).  These contributions involve
either one square (or hexagon) near an end point and the other in
the interior of the chain, or two interfering structures in the
interior of the chain. Also $\eta$ comes from configurations in which
all insertions are near an end of the chain.
 
It seems reasonable to assume that the dominant contribution to $\chi$
can therefore be written as
\begin{eqnarray}
\chi = \sum_n r_0^n [ 1 + n \Delta + n^2 \Delta^2 /2 + \dots ]
\rightarrow \sum r^n \ ,
\label{END2} \end{eqnarray}
where $r=r_0[1 + \Delta]$, as we found in Sec. VI.
The analysis of this appendix
indicates that is appropriate to identify $r$
as the renormalized ratio of a geometric series.
                                                                                
One might ask whether this identification is unique.  Could the
result of Eq. (\ref{END2}) arise from a rounded transition for which
\begin{eqnarray}
\chi= {1 \over 2} \sum_n r_0^n [ (1 + \alpha)^n + (1 +\alpha^* )^n]
\end{eqnarray}
for a suitably chosen value of the complex-valued parameter $\alpha$?  This
form of $\chi$ yields (keeping only relevant terms)
\begin{eqnarray}
\chi &=& \sum_n r_0^n [ 1 + n(\alpha+\alpha^*)/2 + {1 \over 4} n^2( \alpha^2 +
{\alpha^*}^2 ) ]  \ .
\end{eqnarray}
For this to be of the form of Eq. (\ref{END2}) we find that 
$\alpha$ must satisfy
\begin{eqnarray}
{1 \over 4} (\alpha^2 + {\alpha^*}^2)  = {1 \over 8} [\alpha + \alpha^*]^2 \ ,
\end{eqnarray}
which implies that $\alpha^*=\alpha$.  Therefore, we cannot have a rounded
transition by having the real valued critical point for the Bethe
lattice split into a complex conjugate pair of critical points
(which would give a Lorentzian susceptibility with a width of order 
$\sigma^{-2}$).  So the form of Eq. (\ref{END2}) is only consistent with
a shifted pole in the susceptibility, as we implicitly assumed in Sec.  VI.

\end{appendix} 


\begin{thebibliography} {99}
\bibitem{privcommBIROLI} G. Biroli has informed one of us (JMS) about numerical results of another two-dimensional model yielding a discontinuous transition.
\end{thebibliography}
\end{document}